\documentclass[11pt]{article}

\usepackage{amsmath,amssymb,amsthm}
\usepackage{mathtools}
\usepackage{graphicx}
\usepackage{subcaption}
\usepackage{booktabs}
\usepackage{amsfonts}
\usepackage{color}
\usepackage{mathrsfs}
\usepackage{multirow}
\usepackage{rotating}
\usepackage{setspace}
\usepackage{lineno}
\usepackage{url}
\usepackage{float}
\allowdisplaybreaks
\usepackage[colorlinks=true,linkcolor=blue,citecolor=blue]{hyperref}%
\usepackage{array}
\newcolumntype{C}[1]{>{\centering\arraybackslash}m{#1}}
\newtheorem*{theorem*}{Theorem}
\newtheorem{theorem}{Theorem}[section]

\newtheorem{lemma}[theorem]{Lemma}
\newtheorem{proposition}{Proposition}

\theoremstyle{definition}

\newtheorem{remark}{Remark}

\theoremstyle{definition}

\usepackage{geometry}
\numberwithin{equation}{section}
\begin{document}
\title{A novel ``minimal'' model for additional food mediated biological control}

\author{
Kanishka Goyal$^{1}$,
Chanaka Kottegoda$^{2}$,
and Rana D. Parshad$^{1}$
\\[1ex]
\small $^{1}$Department of Mathematics, Iowa State University,
Ames, Iowa 50011, USA
\\
\small $^{2}$Department of Mathematics and Physics, Marshall University,
Huntington, West Virginia 25755, USA
\\[1ex]
\small \texttt{\href{mailto:kgoyal@iastate.edu}{kgoyal@iastate.edu}},
\texttt{\href{mailto:kottegoda@marshall.edu}{kottegoda@marshall.edu}},
\texttt{\href{mailto:rparshad@iastate.edu}{rparshad@iastate.edu}}
}
\maketitle
\begin{abstract}
    Additional food sources are known to influence the dynamics of predator-prey systems, with important implications for biological control. Previous studies of additional food models have incorporated a slew of other factors, such as  predator competition, harvesting and Allee effects. In this work, we propose a \emph{minimal} additional food model, that has features of all of the above mechanisms. Here, we establish sufficient conditions for finite-time predator extinction, and unbounded predator growth, and show that this system can admit up to three interior equilibria. Furthermore, we demonstrate the existence of a Bogdanov-Takens bifurcation of codimension-3, together with the degenerate Hopf bifurcation of codimension at least 2 and cusp bifurcation. Numerical investigations show complex basin structure where distinct long-term outcomes can occur depending on the initial population densities. We discuss applications of our results to pest management strategies.     
\end{abstract}

  %Additional food sources are known to influence the dynamics of predator-prey systems, with important implications for biological control.  nonlinear predator mortality may substantially alter population dynamics. Previous studies of additional food models incorporating generalized predator competition, characterized by $(1<p\leq2)$, revealed a rich bifurcation structure, including a cusp-type degenerate Bogdanov-Takens bifurcation of codimension at least 4, Hopf and homoclinic bifurcations of codimension-3. Also, prey extinction state was achievable. In this work, we investigate the additional food model with sublinear predator mortality, characterized by $(0<p<1)$, and the Holling type-II functional response. Here, we establish sufficient conditions for finite-time predator extinction, and unbounded predator growth, and show that this system can admit up to three interior equilibria. Furthermore, we demonstrate the existence of a Bogdanov-Takens bifurcation of codimension-3, together with the degenerate Hopf bifurcation of codimension at least 2 and cusp bifurcation. Numerical investigations show complex basin structure where distinct long-term outcomes can occur depending on the initial population densities.


\section{Introduction}\label{sec:intro}
Invasive species continue to cause substantial agricultural losses worldwide \cite{roy2024curbing, PSCEWT16, PZM05}. Examples of these in the North-Central United States are, the European corn borer, Western corn rootworm, and soybean aphid, all chief pests on corn and soybean crop. These cause damage of more than $\$3$ billion in crop yield annually \cite{TS14, CO18, BOP22}. Conventional pest control methods rely on pesticides, despite their harmful environmental effects \cite{PB14}. To address this issue, top-down biological control - which is the introduction of natural enemies like predators and/or parasitoids to prey upon the pest population \cite{V96, S99}, has provided a sustainable and environmentally friendly alternative. When predators alone are insufficient for pest control, providing them with additional food (AF) can improve their effectiveness \cite{SV06, T15}. To this end, a number of biological control models have been developed that describe predator-pest dynamics in the presence of AF \cite{SP07, SP10, S02, SPD17, SPV18, VA22, verma2026t}.

Soybean aphid, \textit{Aphis glycines} Matsumura (Hemiptera: Aphididae) is a prominent invasive pest in the North-Central United States, native to Asia and first noticed in southeastern Wisconsin in July 2000 \cite{ragsdale2004soybean}. Its ability to reproduce rapidly, disperse over long distances and favorable environmental conditions have allowed the soybean aphid to quickly establish itself as a major pest \cite{rhainds2008toward}. It has a heteroecious (needs two different host plants for its development) life cycle, considering common Buckthorn, \textit{(Rhamnus cathartica L.)}, as its primary host and soybean, \textit{Glycine max (L.) Merr.}, the secondary host \cite{tilmon2011biology}. During the growing season, reproduction of aphids happen parthenogenetically (reproduction without males) on soybean, which increases their population rapidly, while in the fall, they return to buckthorn for mating and overwintering. They feed on phloem and can colonize almost all above-ground parts of the soybean plant. Heavy infestations can result reduction in yield of up to 40\% \cite{song2006profitability,kim2008economic}, making effective management essential. Soybeans in North America were limitedly treated with insecticides before the invasion of soybean aphid in the continent. As they got established, insecticide use has increased about a 130-fold \cite{ragsdale2011ecology}. Chemical insecticides like pyrethroids and organophosphates are used widely to control these aphids, despite several harmful effects \cite{bahlai2010choosing,desneux2004effects,desneux2007sublethal,varenhorst2012response}. Moreover, they can lead to resistance in soybean aphid populations, against the treatment \cite{hanson2017evidence}. These limitations have led to increasing interest in integrated pest management (IPM) strategies combining cultural practices, aphid-resistant cultivars, and biological control \cite{rutledge2004soybean,johnson2009probability}.

Several studies have investigated biological control models that incorporate additional food (AF) into classical predator-prey models after recognizing the potential benefits of additional food in biological control practices \cite{van2012biological}. In particular, Srinivasu and coauthors \cite{SP07,SP10} developed a predator-prey model in which an introduced predator population $y(t)$ feeds on a target pest population $x(t)$ while also receiving AF. The model is formulated as
\begin{equation}
\label{Eq:1}
\frac{dx}{dt} = x\left(1-\frac{x}{K}\right) - \frac{xy}{1+\alpha \xi + x}, \quad    \frac{dy}{dt} =\frac{\beta\, xy}{1+\alpha \xi + x} + \frac{\beta\, \xi\, y}{1+\alpha \xi + x} - \delta y.
\end{equation}
Here, all parameters are positive, with $K$, $\beta$, and $\delta$ representing the prey carrying capacity, predator conversion efficiency, and natural predator mortality rate, respectively. The parameters $\xi$ and $1/\alpha$ characterize the quantity and quality of AF supplied to the predator. The term $x/(1+\alpha\xi+x)$ represents the Holling type II functional response modified by AF, while $\beta(x+\xi)/(1+\alpha\xi+x)$ gives the corresponding per capita predator growth response. When $\xi=0$, that is there is no AF, the model reduces to a classical Gause type model with type II response. Note, Gause-type predator-prey systems have gained extensive attention in the literature with respect to different ecological applications and functional responses \cite{mukherjee2020bifurcation,huang2013bifurcation,huang2014bifurcations}. Depending on the complexity of the model, the functional response and the ecological mechanism in consideration, these models can exhibit complex dynamics, multiple equilibria, periodic oscillations, and local and global bifurcations. Particularly, models with Holling type II, III, and IV functional responses have been investigated for saddle-node, Hopf, homoclinic, and Bogdanov-Takens (BT) bifurcations, including higher-codimension cases (see Table~\ref{Tab:1}).

Studies of system \eqref{Eq:1} have shown that the quantity and quality of AF strongly influences the biological control outcomes which includes predator-prey coexistence, oscillatory dynamics due to Hopf bifurcation, and pest suppression \cite{SP07, SP10}. Subsequent studies considered AF models with other functional responses, including Type III and Type IV, and ecological applications, revealing diverse dynamical behaviors and bifurcation structures \cite{VA22, W23}. Other ecological mechanisms, including harvesting, prey refuge, predator interference, and Allee effects, have also been widely studied and shown to produce rich dynamical behavior and bifurcation structures \cite{huang2014bifurcations,xiao2007multiple,lamontagne2008bifurcation,etoua2010bifurcation,arsie2022predator,arsie2023high}. More recently, Verma \textit{et al.} \cite{verma2026t} investigated a two-patch AF model incorporating predator dispersal and drift, and the findings showed that spatial heterogeneity can influence pest suppression and generate complex dynamics, including periodic oscillations and chaos. 

Although, AF can improve biological control, an important concern is that under favorable conditions, the introduced predators may grow excessively which leads to unintended ecological consequences \cite{SV06,PQB16}. To regulate predators, intraspecific competition can act as a natural mechanism. The inclusion of predator competition in predator-prey systems, commonly studied through Bazykin-type models \cite{deangelis1975model, bazykin1998nonlinear}, further enriches the dynamics and their global bifurcation structures were recently studied \cite{lu2021global}. Motivated by this, Parshad \textit{et al.} \cite{parshad2023additional} extended the classical AF model by incorporating predator competition through the term $-c\xi y^2$. Their analysis showed rich bifurcation structures numerically summarized in the Table~\ref{Tab:1}. Although their numerical results suggested the existence of higher-order BT bifurcation, a rigorous investigation remained open. These studies then led to the generalization of the classical quadratic competition term to $-c\xi y^p$, with $1<p\leq2$ that encompasses a broader range of nonlinear predator interactions, including nonlinear harvesting, hyperbolic mortality, and generalized competition \cite{sambath2016stability, antwi2020dynamics, barman2023two}. In our more recent work, we investigated a generalized predator competition model in the single-patch setting with $\delta=0$ and theoretically established the existence of BT bifurcation of co-dimension 2 \cite{verma2026additional}. More recently, the model dynamics was studied in detail where the analysis revealed that the system exhibits several higher-order bifurcations, including a cusp-type (or focus-type) BT bifurcation of codimension at least $4$ (or at most $3$), a global Hopf bifurcation of codimension $3$, and a homoclinic bifurcation of codimension $3$ \cite{goyal2026additional}, which demonstrated that AF can enhance the bifurcation structure of predator competition models compared to competition model without AF \cite{lu2021global}. These results motivated us for further investigation of a nonlinear degradation term resulting in predator loss, beyond $1<p\leq2$ regime - particularly in the sublinear regime, that is when $0<p<1$.
In these cases the $-y^{p}$ is non-smooth, posing various hindrances in
the classical mathematical analysis of these systems. For example, linearization about the
trivial steady state is no longer possible \cite{sugie1997predator}, due to the singular nature of the extinction equilibrium. Note, in these
cases one might ask what actual ecological scenarios do these models represent. The
work by Sugie \cite{sugie1997predator}, proposed such terms to be indicative of a highly eﬃcient predator, with a high attack rate.

\begin{table}
\caption{Comparison of bifurcations and dynamics in predator-prey models with/without AF.}
\label{Tab:1}
\centering
{%
\begin{tabular}{|c|C{3cm}|C{3.5cm}|C{5cm}|C{3cm}|}
\hline
& \multicolumn{2}{c|}{Without AF} & \multicolumn{2}{c|}{With AF} \\
\hline
& Functional form & Dynamics studied & Functional form & Dynamics studied \\
\hline\hline

\multirow{3}{*}{(i)} & \multirow{3}{2.5cm}{\centering $f(x) = \frac{x}{1+ x}$}
& \multirow{3}{3cm}{\centering Uniqueness, global stability of limit cycles}
&\vspace{2mm} $f(x,\xi,\alpha) = \frac{x}{1+\alpha \xi + x}$
& \multirow{3}{3cm}{\centering Saddle-node, Hopf,  Transcritical} \\
& 
& 
& \vspace{2mm} $g(x,\xi,\alpha) = \frac{\eta (x+\xi)}{1+\alpha \xi + x}$ & \\ & & \cite{cheng1981uniqueness, kuang1988uniqueness}  & & \cite{SP07}  \\ 
\hline

\multirow{2}{*}{(ii)}
& $f(x) = \frac{x^2}{ax^2+bx+ 1}$, &\vspace{1mm} Saddle-node, Nilpotent saddle (codim-3), Hopf (codim-1 \& 2), BT (codim-2 \& 3)
& $f(x,\xi,\alpha) = \frac{x^2}{1+\alpha\xi^2+x^2}$ & Transcritical, Hopf \\
& $f(x)$ $+$ harvesting & \cite{lamontagne2008bifurcation, etoua2010bifurcation} & $g(x,\xi,\alpha) = \frac{\eta (x^{2}+\xi^{2})}{1+\alpha \xi^{2} + x^{2}}$ & \cite{SPV18, ananth2021influence} \\
\hline

\multirow{5}{*}{(iii)}
& \multirow{5}{3cm}{\centering $f(x) = \frac{x}{ax^2+bx+ 1}$} & \multirow{5}{3.5cm}{\centering Saddle-node, Hopf (codim-2), Homoclinic, BT (codim-3)}
& \vspace{5mm} $f(x,\xi,\alpha) = \frac{x}{(1+\alpha \xi)( \omega x^{2}+1) + x}$ & \multirow{5}{3cm}{\centering Transcritical, Hopf}\\ 
&  &  & \vspace{5mm} $g(x,\xi,\alpha) = \frac{\eta (x+\xi(\omega x^{2}+1))}{(1+\alpha \xi)(\omega x^{2} + 1) + x}$ & \\& & & &\\
& & \vspace{2mm} \cite{zhu2003bifurcation, xiao2001global, xiao2006multiple, rothe1992multiple}
& 
& \cite{ananth2021optimal}   \\ 

\hline
\multirow{2}{*}{(iv)}
& $f(x)$ as in (i),
  with predator competition
& Nilpotent cusp BT (codim-2),
  nilpotent focus BT (codim-3),
  Hopf (codim-2)
& $f(x,\xi,\alpha)$ and $g(x,\xi,\alpha)$
  as in (i), with generalized predator competition $c\xi y^p$, $1<p\leq2$
& Codim-1: Saddle-node, Transcritical; Codim-2: SNTC, CTC, PTC; BT (codim$\geq4$); Codim-3: Hopf, homoclinic\\
& & \cite{lu2021global}
& & \cite{parshad2023additional,verma2026additional} \\
\hline

(v)
& $f(x)$ as in (i),
  with nonlinear predator harvesting
  $\dfrac{hy}{h+y}$
& Multiple equilibria, saddle-node,
  transcritical, Hopf, BT,
  and complex global dynamics
  \cite{lv2019bifurcations}
& $f(x,\xi,\alpha)$ and
  $g(x,\xi,\alpha)$ as in (i),
  with sublinear predator loss
  $cy^p$, $0<p<1$
& predator FTE, predator blowup,
  Codim-2: Hopf, cusp, SNTC; Codim-3: BT.
  Present study \\
\hline
\end{tabular}%
}
\end{table}

In classical predator-prey models, predator mortality is commonly assumed to be proportional to predator density, implying constant per capita mortality rate. However, several studies have considered nonlinear predator mortality functions that accounts for effect of predator density on its death rate. For instance, Cavani and Farkas \cite{cavani1994bifurcations} introduced a bounded predator-specific mortality function that increases with predator density and approaches a limiting value. In addition to density-dependent mortality, nonlinear harvesting has also been studied as a mechanism that influences population dynamics of predators. Gupta \textit{et al.} \cite{gupta2015dynamical} studied a predator-prey model with nonlinear predator harvesting and reported richer dynamics, including saddle-node, Hopf, and Bogdanov-Takens bifurcations, compared to the model without harvesting. Similarly, Zhu and Kong \cite{zhu2017bifurcations} investigated nonlinear predator harvesting in a Leslie-Gower model and established the occurrence of higher-codimension bifurcations. More recently, Mondal \textit{et al.} \cite{mondal2025predator} incorporated predator-specific mortality into an AF mediated predator-prey model with prey refuge, while Pal \textit{et al.} \cite{pal2026exploring} examined the effects of nonlinear mortality on the bifurcation structure of predator-prey systems. These studies show that the form of predator mortality or harvesting can significantly influence the population dynamics. Moreover, there are several applications where harvesting and stocking models are being used for current biological control practices, \cite{lyu2020comparison, juan2025evaluating, mccormick2021simulated}. 

The Allee effect can occur whenever fitness of an individual in a small or sparsely distributed population, decreases as the population size or density does \cite{courchamp2008allee}. The analysis, modeling and consequences of this are a central issue in  population and community ecology. The effect is typically modeled as $\frac{du}{dt}=u(u-m)(1-\frac{u}{K})$ - essentially a modification to the logistic equation. Here 
the state variable $u$, represents the numbers/density of a population at a given time. $m$ is the Allee threshold, with 
$m<K$. A strong Allee effect occurs if $m>0$, and populations falling below this threshold will go extinct . If $m<0$, then there is a weak Allee effect, and we have a compensatory growth function \cite{courchamp2008allee}, where growth rate reduces at low densities. Allee effects have been considered in AF models as well \cite{sasmal2017predator} - but since the Allee mechanism is typically modeled as above, incorporating other mechanisms, such as harvesting, will lead to complicated equations, with multiple terms. Also, there is much recent interest in rate induced extinctions, \cite{vanselow2019very}, where extinction can be induced by dangerous levels of environmental change, without a parameter changing, inherent to classical bifurcations. This often times involves multiple time scale analysis \cite{kaklamanos2022bifurcations}. In our setting, the form of the degradation term, $c y^p$, provides this possibility in our case - even though we have an autonomous system.

Motivated by these findings, we consider a sublinear predator degradation term of the form $c y^p$, where $c>0$ and $0<p<1$, which may represent sublinear harvesting or enhanced density-dependent predator mortality. Unlike linear mortality, the corresponding per capita loss $cy^{p-1}$ increases as predator density decreases and becomes unbounded as $y\to0^+$. Hence, the nonlinear loss can dominate predator growth at sufficiently low densities, making finite-time predator extinction possible even in the presence of AF. Under these assumptions, we consider the following model:

\begin{equation} \label{eq:model_BT}
\begin{aligned}
\dot{x}& = x \left (1-\frac{x}{\gamma}\right )-\frac{x y}{1+x+\alpha \xi} \\
\dot{y}& = \beta \left ( \frac{x+\xi}{1+x+\alpha \xi} \right) y -\delta y-c y^p.  \\
\end{aligned}
\end{equation}
The preliminary analysis of the model are provided in Section~\ref{sec:prelim}, and the detailed stability analysis is given in Section~\ref{sec:stab_analysis}, talking about both boundary and interior equilibria. Section~\ref{sec:bifurcation} presents analysis of possible \textit{higher} codimension bifurcations. The analytical results are then validated numerically in Section~\ref{sec:ns}. We also compare the different $p$ regimes corresponding to predator loss and this comparison is done in Section~\ref{sec:compare}. Section~\ref{sec:discussion} summarizes the main findings of the paper along with biological applications.

\begin{figure}
    \centering   \includegraphics[width=0.9\linewidth]{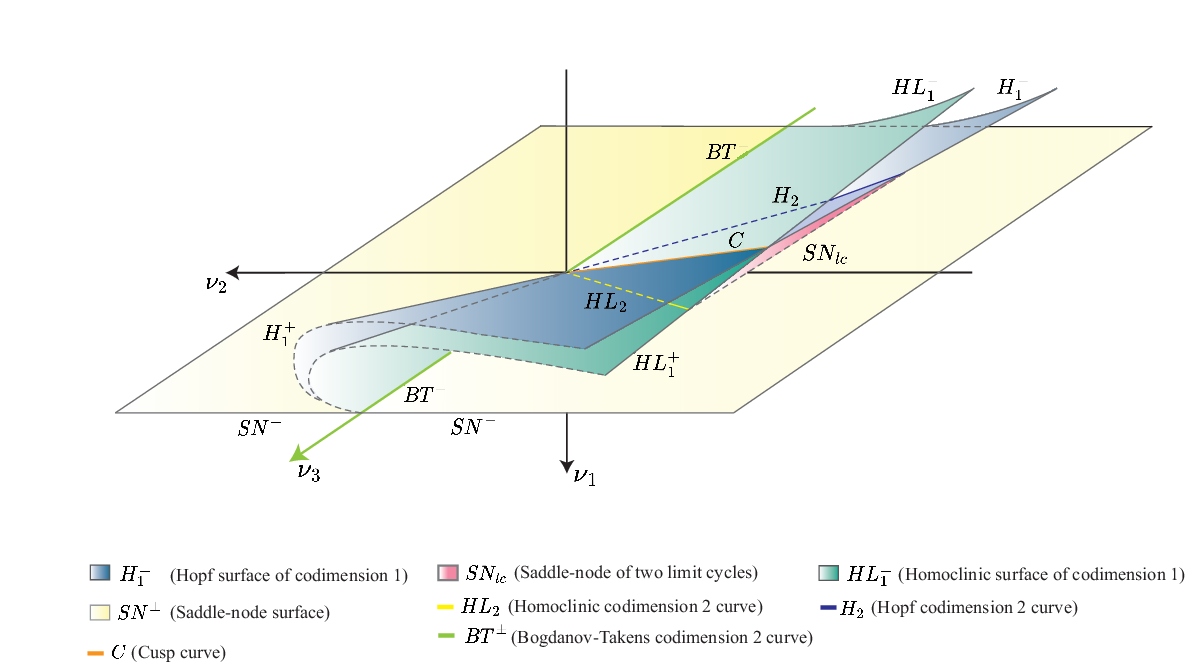}
    \caption{Bifurcation diagram of cusp singularity of codimension 3}
    \label{bt3fig}
\end{figure}

The principal contributions in this paper are as follows:
\begin{itemize}
\item A novel biological control model is introduced with AF for the predator, system~\eqref{eq:model_BT}. This model is \emph{minimal} in the sense, the predator degradation term $-cy^{p}$, can play the role of (i) harvesting (ii) Allee effect (iii) enhanced/hyperbolic predator mortality - and any/all can be chosen depending on the application.

    \item A detailed equilibrium and stability analysis of system~\eqref{eq:model_BT} is performed, establishing the existence and stability of boundary equilibria in Proposition~\ref{prop:boundary_stability} and showing that the system exhibits at most three interior equilibria in Theorem~\ref{thm:max_three_equilibria}.
    \item A sufficient condition under which the predator population undergoes finite-time extinction due to sublinear predator loss is established via Lemma~\ref{lem:FTE}. And the possible unbounded predator growth is established in Lemma~\ref{lem:unbounded_predator}.
    \item We investigate the higher-codimension bifurcations for the system including a codimension-$2$ cusp bifurcation in Theorem~\ref{thm:cusp}.
    \item The existence of codimension-$3$ Bogdanov-Takens bifurcation is proved in Theorem~\ref{thm:BT_existence} and its associated normal form and unfolding is proved in Theorem~\ref{thm:BT_unfolding}. The corresponding dynamics around the BT-point is shown via Figure~\ref{bt3fig}.
    \item Established the existence of codimension-$2$ generalized Hopf (Bautin) bifurcation summarized in Theorem~\ref{thm:gh}.
    \item We construct bifurcation diagram in the $(\delta,\xi)$-plane, identifying regions with different  values and stability types of interior equilibria (see Figure~\ref{fig:region_plots}). We also investigate the interaction between interior equilibria and boundary equilibria, showing the occurence of a saddle-node-transcritical (SNTC) interaction (see Figure~\ref{fig:region_plots}).
    \item Investigated the basins of attraction associated with different long term outcomes and examined how the stable manifolds of saddle equilibria act as separatrix for the regions of finite-time extinction, stable coexistence and unbounded predator growth (see Figure~\ref{fig:FTE_basins}). These results illustrate the sensitivity of predator persistence to the initial population densities.
    \item Compared the qualitative dynamics of different $p$ regimes, highlighting the effects of the nonlinear predator loss term on equilibrium structure and bifurcation behavior in Table~\ref{tab:p_regime_qualitative}-\ref{tab:p_regime_bifurcations} and Figure~\ref{fig:p_regime_comparison}.
    \item Applications of our results to current pest management practices are discussed in Section~\ref{sec:discussion}.
    
\end{itemize}

\section{Preliminary Analysis}\label{sec:prelim}
We first establish some basic properties of system \eqref{eq:model_BT}. Throughout this paper, all parameters are assumed to be positive and \(0<p<1\). The biologically feasible region is
\[
\mathbb{R}_+^2=\{(x,y):x\geq0,\;y\geq0\}.
\]
For any solution initiated in $\mathbb{R}_+^2$, the prey population remains nonnegative. Moreover, if the predator population reaches $y=0$ in finite time, we adopt the biologically relevant continuation $y(t)=0$ thereafter. Thus, solutions are considered in $\mathbb{R}_+^2$.

Since
\[
\dot{x}=x\left(1-\frac{x}{\gamma}\right)-\frac{xy}{1+x+\alpha\xi}
\leq x\left(1-\frac{x}{\gamma}\right),
\]
comparison with the logistic equation gives
\[
0\leq x(t)\leq \max\{x(0),\gamma\},
\qquad
\limsup_{t\to\infty}x(t)\leq\gamma.
\]
Hence, the prey population is ultimately bounded by its carrying capacity $\gamma$.

\begin{remark}
The case $0<p<1$ is qualitatively different: the vector field is not locally Lipschitz at $y=0$, and the nonlinear term $-c y^p$ does not dominate the linear growth term for large $y$. In particular, uniform boundedness of the predator population cannot be concluded from the above argument. However, one can show finite-time extinction of the predator population for sufficiently small initial data [see Lemma~\ref{lem:FTE}.]
\end{remark}

\section{Steady-state analysis}\label{sec:stab_analysis}
We now characterize the steady states of system \eqref{eq:model_BT}. Let $\tilde{E}=(\tilde{x},\tilde{y})$ be an equilibrium. The steady-state conditions are
\begin{equation}\label{eq:equilibrium_conditions}
x\left(1-\frac{x}{\gamma}-\frac{y}{1+x+\alpha\xi}\right)=0,
\qquad
y\left[\beta\left(\frac{x+\xi}{1+x+\alpha\xi}\right)
-\delta-c y^{p-1}\right]=0.
\end{equation}
These equations naturally distinguish equilibria lying on the boundary of $\mathbb{R}_+^2$ from those occurring in its interior.

\subsection{Equilibria on the boundary}
We first characterize the equilibria of system \eqref{eq:model_BT} that lie on the boundary of the biologically feasible region. Equation~\eqref{eq:equilibrium_conditions} yields three possible boundary steady states. These are
\[
\tilde{E}_0=(0,0), \quad \tilde{E}_\gamma=(\gamma,0), \quad \tilde{E}_P=(0,\tilde{y})
\]
where $\tilde{y}=\left(\frac{\beta\xi-\delta(1+\alpha\xi)}{c(1+\alpha\xi)}\right)^{\frac{1}{p-1}}$.

Here, $\tilde{E}_0$ denotes the trivial equilibrium at which both populations are absent, whereas $\tilde{E}_\gamma$ is the predator-free state with the prey population at its carrying capacity. The equilibrium $\tilde{E}_P$ corresponds to the absence of prey while the predator population is maintained through the additional food source. Its biological admissibility is subject to the parameter conditions ensuring that $\tilde{y}$ is real and positive.

To study the behavior of solutions in a neighborhood of these equilibria, we linearize system \eqref{eq:model_BT}. The Jacobian matrix evaluated at an arbitrary point $(x,y)$ is
\begin{equation}
J(x,y)=\begin{bmatrix}1-\dfrac{2x}{\gamma}-\dfrac{(1+\alpha\xi)y}{(1+x+\alpha\xi)^2}
&
-\dfrac{x}{1+x+\alpha\xi}\\[0.3cm]
\dfrac{\beta\left(1+(\alpha-1)\xi\right)y}{(1+x+\alpha\xi)^2}
&
\dfrac{\beta(x+\xi)}{1+x+\alpha\xi}-\delta-cp y^{p-1}
\end{bmatrix}.
\label{general_jacobian_uni}
\end{equation}

For $0<p<1$, the vector field is continuous on $\mathbb{R}_+^2$, but it is not continuously differentiable along $y=0$ due to the term $y^p$. In particular,
\[
\frac{\partial}{\partial y}(y^p)=py^{p-1},
\]
which becomes unbounded as $y\to0^+$. Therefore, the standard Jacobian linearization cannot be directly applied to the equilibria $\tilde{E}_0$ and $\tilde{E}_\gamma$. Their local behavior is instead determined directly from the nonlinear system. The following result summarizes the local behavior.

\begin{proposition}\label{prop:boundary_stability}
For $0<p<1$, the boundary equilibria of system \eqref{eq:model_BT} have the following properties:
\begin{enumerate}
    \item The extinction equilibrium $\tilde{E}_0=(0,0)$ is unstable.
    
    \item The prey-only equilibrium $\tilde{E}_\gamma=(\gamma,0)$ is locally asymptotically stable.
    
    \item If $\beta\xi>\delta(1+\alpha\xi)$, the predator-only equilibrium $\tilde{E}_P=(0,\tilde{y})$ exists and is unstable. More precisely, $\tilde{E}_P$ is a saddle if
    $\tilde{y}>1+\alpha\xi$ and an unstable node if
    $\tilde{y}<1+\alpha\xi$. It is nonhyperbolic when
    $\tilde{y}=1+\alpha\xi$.
\end{enumerate}
\end{proposition}
\begin{proof}
    We consider the three boundary equilibria separately.
    \begin{enumerate}
        \item \textit{Stability of $\tilde{E}_0$.} The $x$-axis is invariant under the flow of system \eqref{eq:model_BT}. Indeed, setting $y=0$ reduces the system to
        \[
        \dot{x}=x\left(1-\frac{x}{\gamma}\right),\qquad \dot y=0.
        \]
        For every $x\in(0,\gamma)$,
        \[
        \dot{x}=x\left(1-\frac{x}{\gamma}\right)>0.
        \]
        Hence, for any sufficiently small $\epsilon>0$, the solution starting from $(\epsilon,0)$ moves away from $\tilde{E}_0$ and approaches $(\gamma,0)$. Since such initial conditions can be chosen arbitrarily close to $\tilde{E}_0$, the equilibrium $\tilde{E}_0$ is unstable.

        \begin{remark}
            Here it is important to consider the exception of initial data on the $y$-axis. That is for small enough solutions starting from $(0,\epsilon)$, they could be attracted to $(0,0)$ in finite time. See Lemma \ref{lem:FTE}.
        \end{remark}

        \item \textit{Stability of $\tilde{E}_\gamma$.} At $\tilde{E}_\gamma=(\gamma,0)$, the prey dynamics along $y=0$ are locally attracting. Moreover, the predator equation can be written as
        \[
        \dot{y}=y^p\left[\left(\beta\frac{x+\xi}{1+x+\alpha\xi}-\delta\right)y^{1-p}-c\right].
        \]
        Since $1-p>0$, we have $y^{1-p}\to0$ as $y\to0^+$. Thus, for $(x,y)$ sufficiently close to $(\gamma,0)$ with $y>0$, $\dot{y}<0$.\\
        
        Furthermore, by Lemma~\ref{lem:FTE}, the predator population reaches zero in finite time. Once $y=0$, the prey dynamics reduce to
        \[
        \dot x=x\left(1-\frac{x}{\gamma}\right),
        \]
        for which $x=\gamma$ is locally asymptotically stable. Therefore, solutions initiated sufficiently close to $\tilde E_\gamma$ approach $(\gamma,0)$, and hence $\tilde E_\gamma$ is locally asymptotically stable.

        \item \textit{Stability of $\tilde{E}_P$.} For $\tilde{E}_P=(0,\tilde{y})$, we have $\tilde{y}>0$, and hence the vector field is continuously differentiable in a neighborhood of $\tilde{E}_P$. The eigenvalues of the Jacobian evaluated at $\tilde{E}_P$ are
        \[
        \lambda_1=1-\frac{\tilde{y}}{1+\alpha\xi},
        \qquad
        \lambda_2=\frac{\beta\xi}{1+\alpha\xi}-\delta-cp\tilde{y}^{p-1}.
        \]
        Using the equilibrium relation
        \[
        \frac{\beta\xi}{1+\alpha\xi}-\delta-c\tilde{y}^{p-1}=0,
        \]
        the second eigenvalue reduces to
        \[
        \lambda_2=(1-p)c\tilde{y}^{p-1}>0.
        \]
        Therefore, $\tilde{E}_P$ is always unstable whenever it exists. Furthermore, the sign of $\lambda_1$ gives the stated saddle, unstable-node, and nonhyperbolic cases.
    \end{enumerate}
\end{proof}

\subsection{Interior Equilibria and Their Multiplicity}
We now investigate the interior equilibria of system
\eqref{eq:model_BT}, with particular emphasis on their number and
multiplicity. These properties play an important role in the
bifurcation analysis developed in the subsequent sections.
From now on we consider $\sigma=1+\alpha\xi$ for convenience and introduce the functions 
\begin{align}\label{eq:func}
r(x)&=\frac{x}{x+\sigma},\quad M(x)=\left(1-\frac{x}{\gamma}\right)(x+\sigma),\quad N(x)=\frac{x+\xi}{x+\sigma}.
\end{align}
which expresses system~\eqref{eq:model_BT} as
\begin{equation}\label{eq:model}
\left\{
\begin{aligned}
\dot{x} &= r(x)[M(x)-y],\\
\dot{y} &= y \bigl[\beta N(x) - \delta - c y^{p-1}\bigr],
\end{aligned}
\right.
\end{equation}
Let $E_i=(x_i,y_i)$ be an arbitrary interior equilibrium. From the first equation of \eqref{eq:model}, we obtain
\[
y_i=M(x_i),
\]
Substitution of $y_i=M(x_i)$ into the second equilibrium equation
reduces the determination of interior equilibria to the scalar
equation
\begin{equation}\label{eq:q_function}
q(x):=\beta N(x)-\delta-c \left(M(x)\right)^{p-1}=0,
\qquad x\in(0,\gamma).
\end{equation}
As a result, each root $x_i\in(0,\gamma)$ of \eqref{eq:q_function} determines a unique interior equilibrium, $E_i=\bigl(x_i,M(x_i)\bigr)$.

For $0<p<1$, one can write the equilibrium equation~\eqref{eq:q_function} as
\begin{equation}\label{eq:q_sublinear}
\beta N(x)-\delta=\frac{c}{(M(x))^{1-p}}.
\end{equation}
In particular, any interior equilibrium $E_i$ must satisfy
\begin{equation}\label{eq:necessary_equilibrium_condition}
\beta N(x_i)-\delta>0.
\end{equation}
Rather than trying to solve~\eqref{eq:q_function} explicitly,
we analyze its geometric structure to determine the maximum number of positive roots. The following result provides an upper bound on the number of interior equilibria admitted by the system.

\begin{figure}
    \centering
    \begin{subfigure}[t]{0.31\textwidth}
        \centering
        \includegraphics[width = \textwidth]{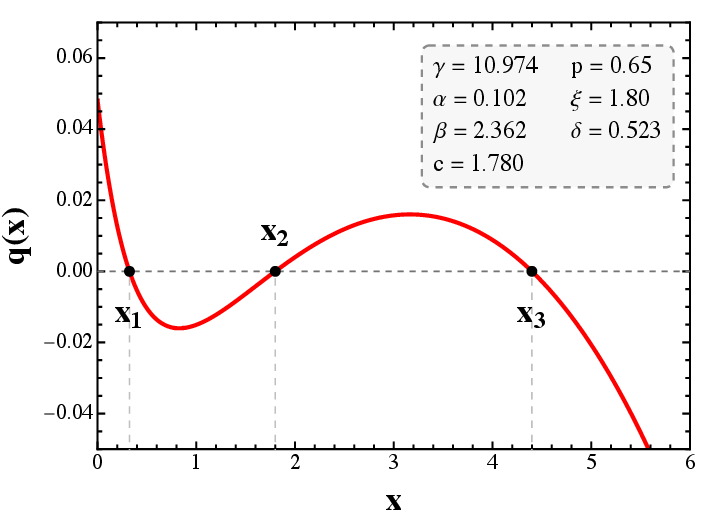}
        \caption{}
        \label{fig:three_roots}
    \end{subfigure}%
    ~ 
    \begin{subfigure}[t]{0.31\textwidth}
        \centering
        \includegraphics[width = \textwidth]{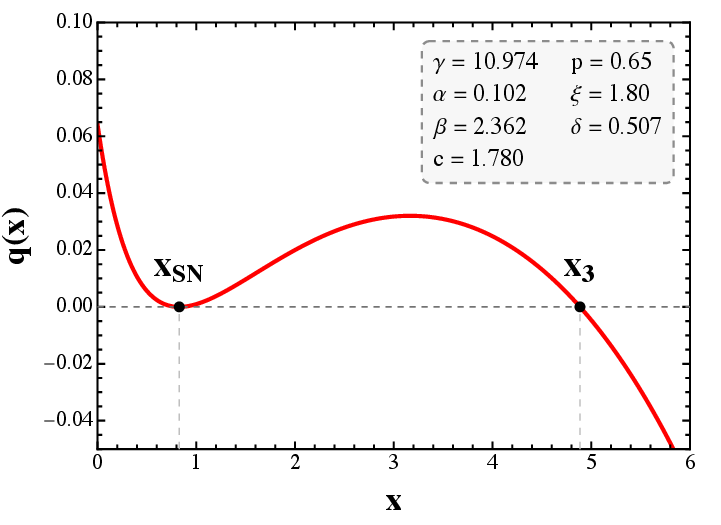}
        \caption{}
        \label{fig:eq_double_single}
    \end{subfigure}%
    ~
    \begin{subfigure}[t]{0.31\textwidth}
        \centering
        \includegraphics[width = \textwidth]{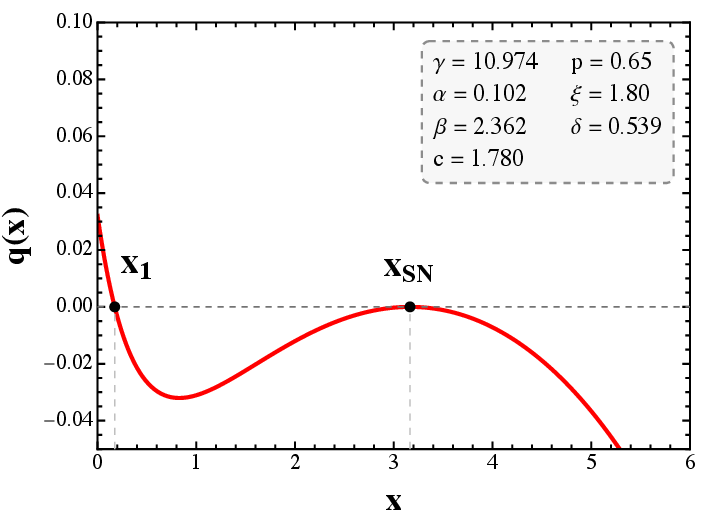}
        \caption{}
        \label{fig:single_double}
    \end{subfigure}\\
    \vspace{1cm}
        \begin{subfigure}[t]{0.31\textwidth}
        \centering
        \includegraphics[width = \textwidth]{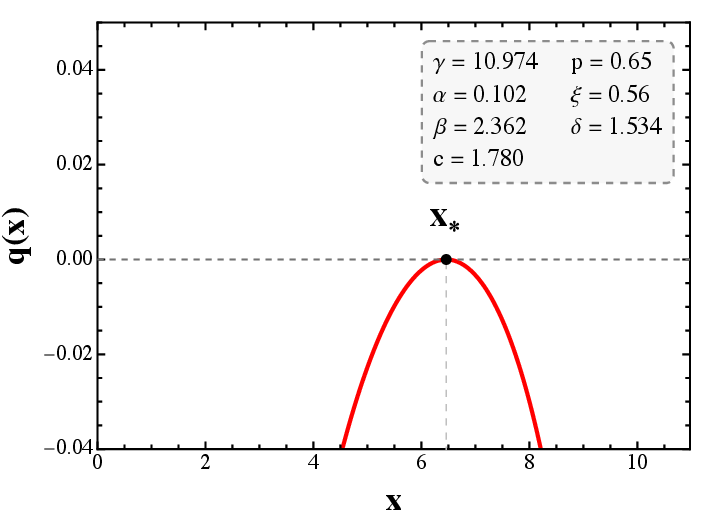}
        \caption{}
        \label{fig:BT}
    \end{subfigure}%
    ~ 
    \begin{subfigure}[t]{0.31\textwidth}
        \centering
        \includegraphics[width = \textwidth]{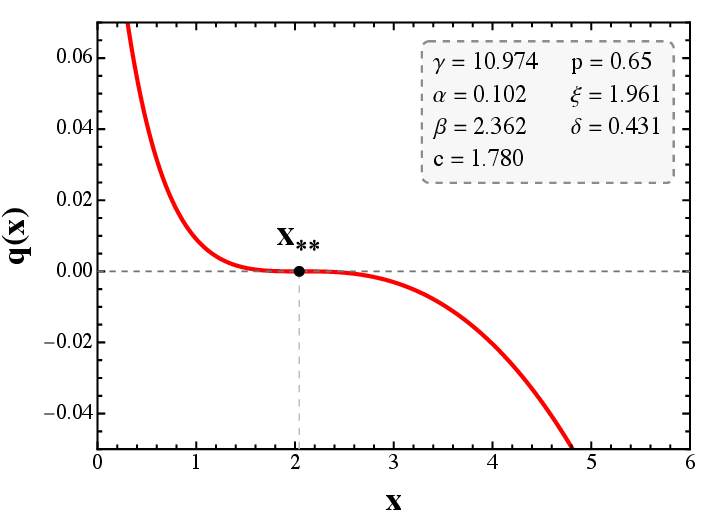}
        \caption{}
        \label{fig:triple_root}
    \end{subfigure}%
    ~
    \begin{subfigure}[t]{0.31\textwidth}
        \centering
        \includegraphics[width = \textwidth]{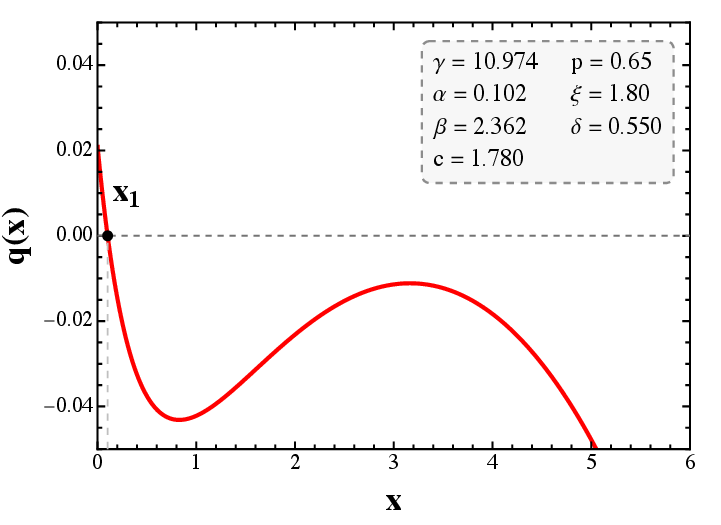}
        \caption{}
        \label{fig:single1}
    \end{subfigure}
    
    \caption{Representative graphs of $q(x)$ illustrating the possible configurations of positive roots of system~\eqref{eq:model_BT} and the corresponding multiplicity of interior equilibria. Panel (a) shows three distinct simple roots $x_1$, $x_2$, and $x_3$; panels (b) and (c) show the coexistence of a simple root and a double root $x_{\mathrm{SN}}$, corresponding to saddle-node bifurcations; panel (d) shows a unique double root $x_*$; panel (e) shows a triple root, $x_{**}$; and panel (f) show a unique simple root $x_1$.}
    \label{fig:root_configurations}
\end{figure}

\begin{theorem}\label{thm:max_three_equilibria}
For $0<p<1$, the system \eqref{eq:model} admits at most
three distinct interior equilibria.
\end{theorem}
\begin{proof}
    First, set $s=1-p\in(0,1)$.
    Equation~\eqref{eq:q_sublinear} can then be written as
    \[
    H(x)=c,
    \]
    where
    \[
    H(x):=\bigl[\beta N(x)-\delta\bigr](M(x))^s.
    \]
    Using the definitions of $M(x)$ and $N(x)$ from~\eqref{eq:func}, gives
    \[
    H(x)=\frac{1}{\gamma^s}\bigl[(\beta-\delta)x+\beta\xi-\delta\sigma\bigr](\gamma-x)^s(x+\sigma)^{s-1}.
    \]
    Since $c>0$, only the interval on which $(\beta-\delta)x+\beta\xi-\delta\sigma>0$ can contain roots of $H(x)=c$.
    Let
    \[
    A=\beta-\delta,
    \qquad
    B=\beta\xi-\delta\sigma.
    \]
    On the admissible interval, $H(x)>0$, and logarithmic differentiation gives
    \[
    \frac{H'(x)}{H(x)}=\frac{A}{Ax+B}-\frac{s}{\gamma-x}+\frac{s-1}{x+\sigma}.
    \]
    Clearly, $H'(x)=0$ if and only if
    \[
    A(\gamma-x)(x+\sigma)-s(Ax+B)(x+\sigma)+(s-1)(Ax+B)(\gamma-x)=0.
    \]
    The left-hand side is a polynomial of degree at most two in $x$. Hence $H(x)$ has at most two critical points in the admissible interval.

    Therefore, by Rolle's theorem, the equation $H(x)=c$ can have at most three distinct roots in $(0,\gamma)$. Since each such root $x_i$ uniquely determines the positive equilibrium 
    \[
    E_i=\bigl(x_i,M(x_i)\bigr), \qquad i=1,\ldots,n,
    \]
    where $n$ denotes the number of distinct roots, we conclude that $n\leq 3$. Hence, system~\eqref{eq:model_BT} admits at most three distinct interior equilibria.
\end{proof}
The possible root configurations of \eqref{eq:q_function} are
illustrated in Figure~\ref{fig:root_configurations}. In particular, the system may exhibit three distinct interior equilibria (Figure~\ref{fig:three_roots}), the coexistence of a simple and a double equilibrium (Figure~\ref{fig:eq_double_single}-\ref{fig:single_double}), a triple equilibrium (Figure~\ref{fig:triple_root}), a unique double root (Figure~\ref{fig:BT}) or a unique simple interior equilibrium (Figure~\ref{fig:single1}), depending on the parameter values as shown in the figure.

The multiplicity of a root of \eqref{eq:q_function} is closely related to the degeneracy of the corresponding interior equilibrium. To establish this connection, the Jacobian matrix of system~\eqref{eq:model} at an interior equilibrium
$E_i=(x_i,M(x_i))$ is given by
\begin{equation}\label{eq:vm}
J(E_i)=
\begin{pmatrix}
r(x_i)M'(x_i) & -r(x_i)\\[1mm]
\beta N'(x_i)M(x_i) & -(p-1)c\bigl(M(x_i)\bigr)^{p-1}
\end{pmatrix},
\end{equation}
The trace and determinant are therefore
\begin{equation}\label{eq:trace_interior}
\operatorname{tr}J(E_i)=r(x_i)M'(x_i)-(p-1)c\bigl(M(x_i)\bigr)^{p-1},
\end{equation}
and
\begin{equation}\label{eq:det_interior}
    \det J(E_i)=r(x_i)M(x_i)\Bigl[\beta N'(x_i)-(p-1)c\bigl(M(x_i)\bigr)^{p-2}M'(x_i)\Bigr].
\end{equation}
The determinant gets simplified to
\begin{equation}\label{eq:det_qprime}
\det J(E_i)=x_i\left(1-\frac{x_i}{\gamma}\right)q'(x_i).
\end{equation}
Since $0<x_i<\gamma$, the factor $x_i(1-x_i/\gamma)$ is strictly positive. Hence the sign of $\det J(E_i)$ is determined by the sign of $q'(x_i)$. In particular,
\begin{itemize}
    \item if $q'(x_i)<0$, then $E_i$ is a hyperbolic saddle;

    \item if $q'(x_i)>0$, then $E_i$ is an elementary
    anti-saddle;

    \item if $q'(x_i)=0$, then $\det J(E_i)=0$, and $E_i$ is a
    degenerate equilibrium.
\end{itemize}
Thus, a multiple root of the scalar equilibrium equation~\eqref{eq:q_function} corresponds to a degenerate interior equilibrium. In particular, a double root satisfies
\begin{equation}\label{eq:double_root_condition}
q(x_i)=q'(x_i)=0,\qquad q''(x_i)\neq0,
\end{equation}
whereas the coalescence of three interior equilibria into a triple root is characterized by
\begin{equation}\label{eq:triple_root_condition}
q(x_i)=q'(x_i)=q''(x_i)=0,\qquad q'''(x_i)\neq 0.
\end{equation}
\begin{remark}
The condition $q'(x_i)=0$ implies
\[
\frac{\beta(\sigma-\xi)}{(x_i+\sigma)^2}=-c(1-p)M(x_i)^{p-2}M'(x_i).
\]
So, a multiple root satisfying $M'(x_i)<0$ can occur only when $\sigma>\xi$. As established in the subsequent bifurcation analysis, $\sigma>\xi$ is a necessary condition for the occurrence of Bogdanov-Takens and Hopf bifurcations.
\end{remark}
\subsection{Finite-Time Extinction and Unbounded Growth}
\begin{lemma}\label{lem:FTE}
    Consider \eqref{eq:model_BT}, with $\alpha \geq 1, \beta > \delta$. Then the predator can go extinct in finite time for sufficiently small predator initial  data.
\end{lemma}

\begin{proof}
    Consider \eqref{eq:model_BT}. By comparison we have,

    \begin{equation} \label{eq:model_BT12}
\begin{aligned}
 \dot{y}& = \beta \left ( \frac{x+\xi}{1+x+\alpha \xi} \right) y -\delta y-c y^p  \\
 & \leq (\beta-\delta)y - c y^{p}.
\end{aligned}
\end{equation}

Standard ODE theory yields, 

\begin{equation}
\lim_{t \rightarrow T^{*}} y(t) \rightarrow 0.
\end{equation}

Thus by comparison we have, for any initial data $y_{0}>0$, such that, $y_{0} < \left(\frac{c}{\beta-\delta}\right)^{\frac{1}{1-p}}=y_c$, $y$ will go extinct in finite time, where an upper bound for the extinction time is, $T^{*} \leq \frac{1}{(\beta-\delta)(1-p)}
\ln{\left(\frac{c}{c - (\beta-\delta)(y_{0})^{1-p}}\right)}$.

\end{proof}

\begin{remark}
    If $\alpha < 1$, an upper estimate of the form,
    $\dot{y} \leq (\beta(\gamma +\xi) - \delta)y - c y^{p}$ works, for $x_{0} < \gamma$ and $\beta(\gamma +\xi) > \delta$. The finite time extinction follows for $(x_{0},y_{0})$ such that, $x_{0} < \gamma$ and $0<y_0<\left(\frac{c}{\beta(\gamma+\xi)-\delta}\right)^{\frac{1}{1-p}}$.
\end{remark}

\begin{remark}
    The term $cy^{p}$ can be seen to play the role of an Allee effect as well. In the absence of the sublinear term, If we think of predator growth rate (at low pest densities), this is given by $\frac{\beta \xi}{1+\alpha \xi} > 0$ - which led to unbounded predator growth \cite{PWB20}. However, with the term in place the predator growth rate (at low pest densities) becomes, $\frac{\beta \xi}{1+\alpha \xi} - cpy^{p-1}$, which could turn negative for predator populations below an Allee threshold, given by $\left(\frac{cp(1+\alpha \xi)}{\beta\xi}\right)^{\frac{1}{1-p}}$. This could be thought of as a strong Allee effect.
\end{remark}

\begin{lemma}\label{lem:unbounded_predator}
    Consider \eqref{eq:model} with $N_m=\min_{0\leq x\leq\gamma}N(x)$. Then the predator population can exhibit infinite-time unbounded growth for sufficiently large initial data.
\end{lemma}
\begin{proof}
    Since $0<x(t)\leq\gamma$, we have $N(x)\geq N_m$. Thus, by comparison we have,
    \begin{equation} \label{eq:model_BT13}
\begin{aligned}
 \dot{y}& =  (\beta N(x) -\delta) y-c y^p  \\
 & \geq (\beta N_m-\delta)y-c y^p.
\end{aligned}
\end{equation}
Let $k=\beta N_m-\delta>0$ and consider
\[
\dot{Y}=kY-c Y^p,\qquad Y(0)=y_0.
\]
Solving this equation gives
\[
Y^{1-p}(t)=\frac{c}{k}+\left(y_0^{1-p}-\frac{c}{k}\right)e^{k(1-p)t}.
\]
Thus, for any initial condition $y_0$ such that $y_0>\left(\frac{c}{k}\right)^{\frac{1}{1-p}}=y_{ug}$, $Y(t)\to\infty$ as $t\to\infty$. Hence, by comparison, $y(t)\geq Y(t)$, and therefore $y(t)\to\infty$ as $t\to\infty$.
\end{proof}
\begin{remark}
    Since $N(x)$ is bounded on $[0,\gamma]$, we have $\dot{y}\leq Cy$ for some $C>0$, which rules out finite-time blow-up. Hence, the unbounded growth occurs only as $t\to\infty$.
\end{remark}

\begin{remark}\label{rem:prey_extinct}
    Since $x(t)$ is bounded, if $y(t)\to\infty$, then $x(t)\to0$ as $t\to\infty$. This follows from
    \[
    \frac{\dot{x}}{x}\leq1-\frac{y(t)}{1+M+\alpha\xi},
    \qquad M=\max\{x_0,\gamma\}.
    \]
    Thus, if $y(t)\to\infty$, the right-hand side becomes negative and is bounded away from zero for sufficiently large $t$. Hence $x(t)\to0$ as $t\to\infty$.
\end{remark}
We now turn to the bifurcation structure of the system and investigate the higher-codimension bifurcations arising from the degeneracy of interior equilibria.
\section{Higher-codimension bifurcations}\label{sec:bifurcation}
\subsection{Bogdanov-Takens Bifurcation: A cusp of codimension-3}
If $E_*$ is a nilpotent singularity corresponding to the double root $x_*$ as in Figure~\ref{fig:BT}, then from the trace and determinant of the variational matrix \eqref{eq:vm}, we get critical values of $\delta$ and $\beta$ as
\begin{equation}
    \begin{aligned}
        \delta&=\frac{x_*(\gamma-\sigma-2x_*)}{\gamma(p-1)(\gamma-x_*)(\sigma-\xi)(\sigma+x_*)}\\&\quad\times
        \Big[\gamma\big(\xi p+(p-1)x_*-\sigma\big)+\xi-p(\xi+x_*)(\sigma+2x_*)+(\xi+2x_*)(\sigma+x_*-1)+2x_*\Big]:=\delta_*,\\
        \beta&=\frac{x_* (2x_*-\gamma+\sigma )^2}{\gamma (\gamma-x_*) (\sigma -\xi )}:=\beta_*>0.
    \end{aligned}
\end{equation}

\begin{remark}\label{remark:delta_positive}
    For $0<p<1$, \(x_*>0\), \(\xi>0\), \(\sigma>0\), and for $\sigma>\xi$, \(\delta_{*}>0\) if
    \[
    x_*<\gamma<\frac{2(p-1)x_*^{2}+(2p-1)x_*\xi+(p-2)x_*\sigma+(p-1)\xi\sigma}{(p-1)x_*+p\xi-\sigma}
    \qquad \text{or} \qquad
    \gamma>2x_*+\sigma.
    \]
\end{remark}
The following theorem establishes the degeneracy of the cusp singularity.
\begin{theorem}\label{thm:BT_existence}
    When $(\delta,\beta)=(\delta_*,\beta_*)$, then the interior equilibrium $E_*$ of system \eqref{eq:model} is a cusp. Moreover,
    \begin{enumerate}
        \item if $\gamma\neq\gamma_*$, then $E_*$ is a cusp of codimension-$2$;
        \item if $\gamma=\gamma_*$, then $E_*$ is a cusp of codimension-$3$,
    \end{enumerate}
    where
    \begin{equation*}
        \gamma_*=\frac{(\sigma +x_*) \sqrt{(12-8 p) x_*^{2}+\sigma ^2+4 \sigma  x_*}+(4 p-6) x_*^2+(2 p-7) \sigma x_*-\sigma ^2}{2 (p-1) x_*-2 \sigma}.
    \end{equation*}
\end{theorem}
\begin{proof}
    When $(\delta,\beta)=(\delta_*,\beta_*)$, we translate the interior equilibrium $E_*$ to the origin and expand system \eqref{eq:model} in a Taylor series around the origin. Let $X=x-x_*$ and $Y=y-M(x_*)$. Then system \eqref{eq:model} can be rewritten as
\begin{eqnarray}\label{eq:s1}
\left\{\begin{aligned}
\dot{X}&= r(x_*)M'(x_*)X -r(x_*)Y +\frac{r(x_*)M''(x_*)+2r'(x_*)M'(x_*)}{2}X^2 -r'(x^*)XY + \mathcal{V}_1(X,Y),\\
\dot{Y}&= r(x_*)(M'(x_*))^2X-r(x_*)M'(x_*)Y+ \frac{\beta M(x_*) N''(x_*)}{2}X^2 +\beta N'(x_*)XY\\
&\quad -\frac{p(p-1)c (M(x_*))^{p-2}}{2}Y^2 +\mathcal{W}_1(X,Y),
\end{aligned}\right.
\end{eqnarray}
where $\mathcal{V}_1$ and $\mathcal{W}_1$ are $C^\infty$ functions of $O(|X,Y|^3)$ in $(X,Y)$.
This system is further transformed to the required codimension 2 normal form by applying the change of variables successively
\begin{equation}
\begin{aligned}
    X&=\bar{x}, \quad \text{and} \quad Y=\frac{1}{r(x_*)}\left(r(x_*)M'(x_*)\bar{x}-\bar{y}\right),\\
    \bar{x}&=\bar{x}+\frac{1}{4} \left(\frac{p M'(x_*)}{M(x_*)}+\frac{2 r'(x_*)}{r(x_*)}\right)\bar{x}^2, \quad \text{and} \quad \bar{y}= \bar{y} -\frac{1}{2}r(x_*)M''(x_*)\bar{x}^2 + \frac{p M'(x_*)}{2M(x_*)}\bar{x}\bar{y},
    \end{aligned}
\end{equation}
With this transformation and using Remark $1$ of section $2.13$ in \cite{perko2013differential}, the system \eqref{eq:s1} is transformed to the standard normal-form:
\begin{eqnarray}\label{eq:s4}
\left\{\begin{aligned}
\dot{\bar{x}}= &\ \bar{y}+ \mathcal{V}_1(\bar{x},\bar{y}),\\
\dot{\bar{y}}=&\ \zeta_1\bar{x}^2 + \zeta_2\bar{x}\bar{y} +\mathcal{W}_1(\bar{x},\bar{y}),
\end{aligned}\right.
\end{eqnarray}
where $\mathcal{V}_1$ and $\mathcal{W}_1$ are $C^\infty$ functions of $O(|X,Y|^3)$ in $(\bar{x},\bar{y})$ and the normal-form coefficients are given by
\begin{align}
    \zeta_1&=\frac{r(x_*)\Bigl(r(x_*)M'(x_*)(pM'(x_*)^2 +M(x_*)M''(x_*))-\beta_*M(x_*)(2M'(x_*)N'(x_*)+M(x_*)N''(x_*))\Bigr)}{2M(x_*)},\\
    \zeta_2&=r'(x_*)M'(x_*)+ \beta_*N'(x_*)+ r(x_*)M''(x_*)-\frac{pr(x_*)M'(x_*)^2}{M(x_*)}.
\end{align}
The coefficients $\zeta_1$ and $\zeta_2$ determine the nondegeneracy of the Bogdanov-Takens singularity. In particular, whenever
\[
\zeta_1\neq 0
\qquad\text{and}\qquad
\zeta_2\neq 0,
\]
the equilibrium $E_*$ is a nondegenerate Bogdanov-Takens singularity of codimension 2 by the Theorem $3$ in Section $2.11$ in \cite{perko2013differential}. Notice that no restriction on the sign of $\zeta_1$ is required; its nonvanishing is sufficient for the quadratic nondegeneracy condition. On the other hand, when $\zeta_2=0$, i.e.,
\begin{equation}
    \gamma=\frac{(x_*+\sigma) \sqrt{(12-8 p) x_*^2+\sigma ^2+4 \sigma  x_*}+(4 p-6) x_*^2+(2 p-7) \sigma  x-\sigma ^2}{2 (p-1) x_*-2 \sigma }:=\gamma_*
\end{equation}
then $E_*$ is a cusp of codimension at least $3$. Under this condition, direct simplification shows that the coefficient $\zeta_1>0$.

Now when $\zeta_2=0$, we investigate the exact codimension of the cusp $E_*$. To do so, we rewrite system \eqref{eq:s4} as the following by applying $C^\infty$ change of coordinates of the form $\bar{x}_1=\bar{x}$ and $\bar{y}_1= \dot{\bar{x}}$

\begin{equation}\label{eq:s5}
\left\{
\begin{aligned}
\dot{\bar{x}}_1 &= \bar{y}_1,\\
\dot{\bar{y}}_1&=\sum_{i=2}^{4} a_{i0}\bar{x}_1^{\,i}+\sum_{\substack{1\leq i,j\leq 3\\ 3\leq i+j\leq 4}}a_{ij}\bar{x}_1^{\,i}\bar{y}_1^{\,j}+\sum_{j=3}^{4} a_{0j}\bar{y}_1^{\,j}+\mathcal{W}_2(\bar{x}_1,\bar{y}_1).
\end{aligned}
\right.
\end{equation}
% \begin{eqnarray}\label{eq:s5}
% \left\{\begin{aligned}
% \dot{\bar{x}}_1= &\ \bar{y}_1,\\
% \dot{\bar{y}}_1=&\ a_{20}\bar{x}_1^2 + a_{30}\bar{x}_1^3 + a_{21}\bar{x}_1^2\bar{y}_1 + a_{12}\bar{x}_1\bar{y}_1^2 + a_{03}\bar{y}_1^3 + a_{40} \bar{x}_1^4 + a_{31}\bar{x}_1^3\bar{y}_1 + a_{13}\bar{x}_1\bar{y}_1^3 + a_{22}\bar{x}_1^2\bar{y}_1^2+a_{04}\bar{y}_1^4+\mathcal{W}_2(\bar{x}_1,\bar{y}_1),
% \end{aligned}\right.
% \end{eqnarray}
where $\mathcal{W}_2$ is $C^\infty$ function of $O(|X,Y|^5)$ in $(\bar{x}_1,\bar{y}_1)$ and the coefficients $a_{i0}$, $a_{ij}$ and $a_{0j}$ are given in Appendix \ref{sec:A1}.\\
 
Further, we apply a near-identity transformation to the system \eqref{eq:s5} of the form 
\begin{equation}\label{eq:t4}
    \bar{x}_1 \;\mapsto\; \bar{x}_1+\bar{P}(\bar{x}_1,\bar{y}_1), 
    \qquad 
    \bar{y}_1 \;\mapsto\; \bar{y}_1+\bar{Q}(\bar{x}_1,\bar{y}_1),
\end{equation}
where $\bar{P}$ and $\bar{Q}$ are smooth polynomials
\begin{eqnarray}\label{functions1}
\begin{aligned}
\bar{P}(\bar{x}_1,\bar{y}_1)&=\left(\frac{a_{03}}{2}\right)\bar{x}_1^2\bar{y}_1 + \left(\frac{a_{04}}{2}\right)\bar{x}_1^2\bar{y}_1^2 + \left(\frac{a_{13}}{6}\right)\bar{x}_1^3\bar{y}_1, \\
\bar{Q}(\bar{x}_1,\bar{y}_1)& = a_{03}\bar{x}_1\bar{y}_1^2+a_{04}\bar{x}_1\bar{y}_1^3+\left(\frac{a_{13}}{2}\right)\bar{x}_1^2\bar{y}_1^2 + \left(\frac{a_{03}a_{20}}{2}\right)\bar{x}_1^4
\end{aligned}
\end{eqnarray}
which reduces system \eqref{eq:s5} to
\begin{equation}\label{eq:s6}
\left\{
\begin{aligned}
\dot{\bar{x}}_1 &= \bar{y}_1,\\
\dot{\bar{y}}_1&=\sum_{i=2}^{4} b_{i0}\bar{x}_1^{\,i}
+\sum_{i=2}^{3} b_{i1}\bar{x}_1^{\,i}\bar{y}_1+\sum_{i=1}^{2} b_{i2}\bar{x}_1^{\,i}\bar{y}_1^{\,2}+\mathcal{W}_2(\bar{x}_1,\bar{y}_1).
\end{aligned}
\right.
\end{equation}
% \begin{eqnarray}\label{eq:s6}
% \left\{\begin{aligned}
% \dot{\bar{x}}_1= &\ \bar{y}_1,\\
% \dot{\bar{y}}_1=&\ b_{20}\bar{x}_1^2 + b_{30}\bar{x}_1^3 + b_{21}\bar{x}_1^2\bar{y}_1 + b_{12}\bar{x}_1\bar{y}_1^2 + b_{40} \bar{x}_1^4 + b_{31}\bar{x}_1^3\bar{y}_1 + b_{22}\bar{x}_1^2\bar{y}_1^2 +\mathcal{W}_2(\bar{x}_1,\bar{y}_1),
% \end{aligned}\right.
% \end{eqnarray}
where $\mathcal{W}_2$ is $C^\infty$ function of $O(|X,Y|^5)$ in $(\bar{x}_1,\bar{y}_1)$ and the coefficients $b_{i0}$, $b_{i1}$ and $b_{i2}$ are given in Appendix \ref{sec:A2}

Make the following change of variables
\begin{align}
    X=\bar{x}_1, \quad Y=\frac{\bar{y}_1}{\sqrt{b_{20}}}, \quad d\tau=\sqrt{b_{20}}dt,
\end{align}
which brings system \eqref{eq:s6} to the form (up to fourth order),
\begin{eqnarray}\label{eq:s7}
\left\{\begin{aligned}
\dot{X}= &\ Y,\\
\dot{Y}=&\ X^2 + C_1X+C_2X^2 Y + C_3XY^2 + C_4X^4 + C_5X^3Y + C_6X^2Y^2 + \mathcal{W}_2(X,Y),\\
\end{aligned}\right.
\end{eqnarray}
where we denote $\tau$ by $t$, $\mathcal{W}_2$ is $C^\infty$ function of $O(|X,Y|^5)$ in $(X,Y)$ and the coefficients are given as
\begin{align*}
C_1&=\frac{b_{30}}{b_{20}}, \quad C_2=\frac{b_{21}}{\sqrt{b_{20}}}, \quad C_3=b_{12}, \quad C_4=\frac{b_{40}}{b_{20}}, \quad C_5=\frac{b_{31}}{\sqrt{b_{20}}}, \quad C_6=b_{22}.
\end{align*}
By the Proposition $5.3$ given in \cite{lamontagne2008bifurcation}, the equivalent system of \eqref{eq:s7} is 
\begin{eqnarray}\label{eq:s8}
\left\{\begin{aligned}
\dot{X}= &\ Y ,\\
\dot{Y}=&\ X^2+\zeta_3X^3Y +\mathcal{W}_2(X,Y),
\end{aligned}\right.
\end{eqnarray}
where $\mathcal{W}_2$ is $C^\infty$ function of $O(|X,Y|^5)$ in $(X,Y)$ and
\begin{align}
\zeta_3&=\frac{b_{20}b_{31}-b_{30}b_{21}}{\left(b_{20}\right)^{3/2}}.
\end{align}
Note that the entire expression for $\zeta_3$ has been omitted here for brevity and using the \textit{Reduce} command in \textit{Mathematica}, it was found that $\zeta_3$ is always positive for the considered biologically relevant parameter restrictions.\\

Therefore, by the normal form theory in \cite{chow1994normal}, we observe that when $\zeta_3\neq0$ the equilibrium point $E_*(x_*,G(x_*))$ behaves as a cusp of codimension-$3$.
\end{proof}

In the following theorem, we discuss if system \eqref{eq:model} can undergo Bogdanov-Taken bifurcation under a small parameter perturbation.
\begin{theorem}\label{thm:BT_unfolding}
    Suppose that $(\beta,\delta,\gamma)=(\beta_*,\delta_*,\gamma_*)$. Then system \eqref{eq:model} admits a cusp singularity $E_*$ of codimension-3. If we choose $\delta$, $\beta$ and $\gamma$ as the bifurcating parameters then the system undergoes Bogdanov-Taken bifurcation in a sufficiently small neighborhood of $E_*$ as $(\beta,\delta,\gamma)$ varies near $(\beta_*,\delta_*,\gamma_*)$. 
\end{theorem}

\begin{proof}
    We take $(\beta,\delta,\gamma)$ as the bifurcation parameters and introduce the perturbations
    \[
    \beta=\beta_* +\epsilon_1,\qquad
    \delta=\delta_* +\epsilon_2,\qquad
    \gamma=\gamma_* +\epsilon_3.
    \]
    Accordingly, we consider the perturbed system
\begin{equation}\label{eq:perturbed_system}
\left\{\begin{array}{l}
\displaystyle \dot{x}=x \left (1-\frac{x}{\gamma_* + \epsilon_3}\right )-\frac{x y}{1+x+\alpha\xi} ,\\
\displaystyle \dot{y}=y\,\left[ (\beta_* + \epsilon_1)\left ( \frac{x+\xi}{1+x+\alpha\xi} \right)\ -(\delta_* + \epsilon_2)\, - c y^{(p-1)}\right].
\end{array}\right.
\end{equation}
   We transform system \eqref{eq:perturbed_system} into its universal unfolding through the following sequence of coordinate transformations and time rescalings.\\
\medskip
\noindent\textbf{Step 1.}
   Starting with the translation of the equilibrium point $E_*$ to the origin by introducing the change of variables
    \[
    (u, v) \;\mapsto\; \left(x-x_*,\, y-y_*\right)
    \]
    Expanding the resulting system in a Taylor series about the origin and retaining terms up to fourth degree, with the remaining terms absorbed into the higher-order remainder, yields
\begin{equation}\label{eq:unfolding_system1} 
\left\{ \begin{aligned} 
\dot{u} &= \sum_{i=0}^{4}c_{i0}u^{i} +v\sum_{i=0}^{3}c_{i1}u^{i} +O(|u,v,\epsilon|^5),\\[1ex] 
\dot{v} &=\sum_{i=0}^{4}d_{i0}u^{i} +v\sum_{i=0}^{3}d_{i1}u^{i} +\sum_{j=2}^{4}d_{0j}v^{j} +O(|u,v,\epsilon|^5), 
\end{aligned} \right.
\end{equation}
The coefficients $c_{ij}$ and $d_{ij}$ are given in Appendix~\ref{sec:A3}. \\
\medskip
\noindent\textbf{Step 2.} Set $X=u$ and $Y=\dot{u}$. Then system \eqref{eq:unfolding_system1} can be written as
\begin{equation}\label{eq:unfolding_system2} 
\left\{ 
\begin{aligned} \dot{X} &= Y,\\[1ex] \dot{Y} &= \sum_{i=0}^{4} e_{i0}X^{i} +Y\sum_{i=0}^{3} e_{i1}X^{i}+Y^{2}\sum_{i=0}^{2} e_{i2}X^{i} + Y^{3}\sum_{i=0}^{1} e_{i3}X^{i}+e_{04}Y^4 + O(|X,Y,\epsilon|^5). 
\end{aligned} \right. 
\end{equation}
Here, the coefficients $e_{ij}$ are expressed in terms of $c_{ij}$ and $d_{ij}$ and their explicit expressions are provided in Appendix~\ref{sec:A3}. \\
\medskip
\noindent\textbf{Step 3.} We next apply the near-identity transformation
\begin{equation}\label{eq:unfolding_transf_2}
    X=X_1+\Phi(X_1,Y_1),\qquad Y=Y_1+\Psi(X_1,Y_1),
\end{equation}
where the functions $\Phi$ and $\Psi$ are given in Appendix~\ref{sec:A3}. Under this transformation, system \eqref{eq:unfolding_system2} takes the form
\begin{equation}\label{eq:unfolding_system3} 
\left\{ 
\begin{aligned} 
\dot{X}_1 &= Y_1,\\[1ex] \dot{Y}_1 &= \sum_{i=0}^{4} f_{i0}X_1^{\,i} +Y_1\sum_{i=0}^{3} f_{i1}X_1^{\,i} +O(|X_1,Y_1,\epsilon|^5). 
\end{aligned} \right. 
\end{equation}
The coefficients $f_{ij}$ are listed in Appendix~\ref{sec:A3}. 

\medskip
\noindent\textbf{Step 4.} We now eliminate the pure $X_1^i$ ($i=3,4$) terms. Since $f_{20}(0)=\zeta_1>0$, we use the differential relation $\sum_{i=2}^{4}f_{i0}X_1^idX_1=f_{20}X_2^2dX_2$ which gives the following 
\begin{equation}
    \begin{aligned}
    X_1=\mathcal{X}(X_2)&:=X_2-\frac{f_{30}}{4f_{20}}X_2^2+\frac{15f_{30}^2-16f_{20}f_{40}}{80f_{20}^2}X_2^3\\
\end{aligned}
\end{equation}
Together with the coordinate transformation and time rescaling
\begin{equation}
\begin{aligned}
X_1 &= \mathcal{X}(X_2), \qquad Y_1 = Y_2, \quad
d\tau = \Bigl(
1 +\tfrac{f_{30}}{2\,f_{20}}\,X_2+\left(\tfrac{3\,f_{40}}{5\,f_{20}} - \tfrac{5\,f_{30}^{2}}{16\,f_{20}^{2}} \right)X_2^{2} + \Bigl( -\tfrac{7\,f_{30}^{3}}{16\,f_{20}^{3}} + \tfrac{3\,f_{30}f_{40}}{5\,f_{20}^{2}}  \Bigr) X_2^{3} \Bigr)dt.
\end{aligned}
\end{equation}
system \eqref{eq:unfolding_system3} reduces to
\begin{equation}\label{eq:unfolding_system4} 
\left\{ 
\begin{aligned} 
\dot{X}_2 &= Y_2,\\[1ex] 
\dot{Y}_2 &= \sum_{i=0}^{4} g_{i0}X_2^{\,i} +Y_2\sum_{i=0}^{3} g_{i1}X_2^{\,i} +R(X_2,Y_2,\epsilon). 
\end{aligned} \right. 
\end{equation}
In particular, $g_{30}(0)=g_{40}(0)=0$. The remaining coefficients $g_{ij}$ are given in Appendix~\ref{sec:A3}, while $R(X_2,Y_2,\epsilon)$ collects the corresponding higher-order terms.

\medskip
\noindent\textbf{Step 5.} Since $g_{20}(0)=\zeta_1>0$, we eliminate the linear $X_2$ term and normalize the coefficient of the quadratic term by applying
\begin{equation}
    \begin{aligned}
    X_2&=X_3-\frac{g_{10}}{2g_{20}}, \quad Y_2=\sqrt{g_{20}}Y_3,\quad \tau= \sqrt{g_{20}}t,
\end{aligned}
\end{equation}
This transforms system~\eqref{eq:unfolding_system4} into
\begin{equation}\label{eq:unfolding_system5}
\left\{\begin{array}{ll}
\displaystyle \dot{X_3} &= Y_3, \\
\displaystyle \dot{Y_3} &=h_{00}+h_{01}Y_3+X_3^2+h_{11}X_3Y_3+h_{30}X_3^3+h_{21}X_3^2Y_3+h_{40}X_3^4 +h_{31}X_3^3Y_3 + R(X_3,Y_3,\epsilon),\\
\end{array}\right.
\end{equation}
We then eliminate the $X_3^2Y_3$-term by means of the transformation
\begin{equation}
    \begin{aligned}
X_3&=X_4, \quad Y_3=Y_4+\frac{h_{21}}{3}Y_4^2+\frac{2h_{21}^2}{9}Y_4^3+\frac{5h_{21}^3}{27}Y_4^4 ,\quad d\tau= \Bigl(1+\frac{h_{21}}{3}Y_4+\frac{2h_{21}^2}{9}Y_4^2+\frac{5h_{21}^3}{27}Y_4^3\Bigr) \, dt.
\end{aligned}
\end{equation}
Consequently, we obtain
\begin{equation}\label{eq:unfolding_system6}
\left\{\begin{array}{ll}
\displaystyle \dot{X_4} &= Y_4, \\
\displaystyle \dot{Y_4} &=J_{00}+J_{01}Y_4+X_4^2+J_{11}X_4Y_4+J_{31}X_4^3Y_4 + R(X_4,Y_4,\epsilon),\\
\end{array}\right.
\end{equation}
where coefficients, $J_{ij}$ are in Appendix~\ref{sec:A3}.

\medskip
\noindent\textbf{Step 6.} Finally, since $J_{31}(0)=\rho>0$, we introduce the scaling
\begin{equation}
    \begin{aligned}
    X_4&=J_{31}^{-\frac{2}{5}}x, \quad Y_4=J_{31}^{-\frac{3}{5}}y,\quad d\tau =J_{31}^{-\frac{1}{5}} dt.
\end{aligned}
\end{equation}
Under this transformation, system \eqref{eq:unfolding_system6} takes the normal form
\begin{equation}\label{eq:unfolding_system7}
\left\{\begin{array}{ll}
\displaystyle \dot{x} &= y, \\
\displaystyle \dot{y} &=\nu_1+\nu_2y+\nu_3xy+x^2+x^3y + R(x,y,\epsilon),\\
\end{array}\right.
\end{equation}
where \(\nu_0,\nu_1,\nu_2\) are smooth functions of the perturbation parameters and their explicit expressions are given in Appendix~\ref{sec:A3}.
\begin{equation} \label{eq:jac_cond} 
\left.\frac{\partial(\nu_1,\nu_2,\nu_3)}
{\partial(\epsilon_1,\epsilon_2,\epsilon_3)}\right|_{\epsilon=0}
=-\frac{(\sigma-\xi)\rho^{4/5}\zeta_1^{-3}x_*^4}{4\mathcal{F}_2\gamma_*^6(\gamma_*-x_*)(x_*+\sigma)^7}\mathcal{F}_1\mathcal{P}(x_*)<0.
\end{equation}
Here, \(\mathcal{F}_1\), \(\mathcal{F}_2\), and the polynomial
\(
\mathcal{P}(x_*)=\sum_{j=0}^{7}\mathcal{P}_jx_*^j
\)
are given explicitly in Appendix~\ref{sec:A3}. Under the critical values of \(\beta,\delta,\gamma\)
\[
\mathcal{F}_1<0,\qquad
\mathcal{F}_2<0,\qquad
\mathcal{P}(x_*)>0,
\]
Thus, the above Jacobian is nonsingular at \(\epsilon=0\). Hence, system \eqref{eq:unfolding_system7} provides a universal unfolding of the cusp singularity of order 3. Consequently, the bifurcation structure of system \eqref{eq:perturbed_system} in the parameter space \(\epsilon_1,\epsilon_2,\epsilon_3\) is locally equivalent, up to a homeomorphism of the parameter space, to that of system \eqref{eq:unfolding_system7} with respect to \(\nu_1,\nu_2,\nu_3\) \cite{chow1994normal}.  
\end{proof}

\noindent
It is clear that system \eqref{eq:unfolding_system7} has no equilibrium points when \(\nu_1>0\) and equilibrium points exists only when \(\nu_1<0\). Therefore, the surface of saddle-node bifurcation of equilibrium points is given by the plane \(\nu_1=0\). The saddle-node can be attracting \((SN^-)\) or repelling \((SN^+)\). Three bifurcation surfaces exist in the half-space \(\nu_1<0\);

\begin{itemize}

    \item \textbf{Hopf bifurcation surface.}
    The surfaces \(H_1^+\) and \(H_1^-\) correspond to codimension-one Hopf bifurcations. In particular, \(H_1^+\) denotes the subcritical Hopf bifurcation surface, whereas \(H_1^-\) denotes the supercritical Hopf bifurcation surface. Along the curve \(H_2\), the system exhibits a codimension-two Hopf bifurcation.

    \item \textbf{Homoclinic bifurcation surface.}
    The surfaces \(HL_1^-\) and \(HL_1^+\) correspond to codimension-one homoclinic bifurcations. Specifically, \(HL_1^-\) represents the occurrence of an attracting homoclinic loop, while \(HL_1^+\) represents a repelling homoclinic loop. Along the curve \(HL_2\), the system undergoes a codimension-two homoclinic bifurcation.

    \item \textbf{Saddle-node bifurcation of limit cycles.}
    The saddle-node bifurcation of limit cycles is denoted by \(SN_{lc}\). Along the surface \(SN_{lc}\), a unique semistable limit cycle exists. Moreover, the surface \(SN_{lc}\) is tangent to the Hopf bifurcation surface \(H_1^-\) along the curve \(H_2\). It is also tangent to the homoclinic bifurcation surface \(HL_1^+\) along the corresponding codimension-two bifurcation curve.

\end{itemize}

 Inside the region bounded by \((HL_1^+)\), \((H_1^-)\) and \((SN_{lc})\), system \eqref{eq:unfolding_system7} has exactly two hyperbolic limit cycles, where the inner one is unstable, and the outer one is stable. At point \(C\), a stable homoclinic loop coexists with a weak focus of order 1 whose first focus quantity is positive \cite{arsie2022predator, arsie2023high}.

\subsection{Hopf Bifurcation}
\subsubsection{Existence of Hopf bifurcation}
Let $E_i=(x_i,y_i)$ be an arbitrary interior equilibrium of system~\eqref{eq:model}, where $y_i=M(x_i)$ and $q(x_i)=0$. We investigate the conditions under which system~\eqref{eq:model} undergoes a Hopf bifurcation at $E_i$. We choose $\delta$ as the bifurcation parameter.

At an interior equilibrium,
\[
\operatorname{tr}J(E_i)=r(x_i)M'(x_i)-(p-1)\bigl(\beta N(x_i)-\delta\bigr).
\]
Thus, the Jacobian has zero trace when
\begin{equation}\label{eq:delta_H}
\delta=\beta N(x_i)-\frac{r(x_i)M'(x_i)}{p-1}=: \delta_H.
\end{equation}
Since the interior equilibrium satisfies
\begin{equation}
q(x_i,\delta)=0,
\end{equation}
implicit differentiation with respect to $\delta$ gives
\[
\frac{dx_i}{d\delta}=\frac{1}{q_x(x_i,\delta)},
\]
where
\[
q_x(x_i,\delta)=\beta N'(x_i)-(p-1)\bigl(\beta N(x_i)-\delta\bigr)\frac{M'(x_i)}{M(x_i)}.
\]
Notice that
\begin{align}
    \det J(E_i)&=r(x_i)M(x_i)\left[\beta N'(x_i)-(p-1)\bigl(\beta N(x_i)-\delta\bigr)\frac{M'(x_i)}{M(x_i)}\right]\\
    &=r(x_i)M(x_i)q_x(x_i,\delta).
\end{align}
Hence, if $q_x(x_i,\delta)>0$, then $\det J(E_i)>0$, and the Jacobian at $\delta=\delta_H$ has a pair of purely imaginary eigenvalues.

We next verify the transversality condition. Let $\lambda(\delta)$ and $\overline{\lambda}(\delta)$ denote the two complex conjugate eigenvalues of $J(E_i)$. 

At $\delta=\delta_H$, we have $\operatorname{tr}J(E_i)=0$ and
$\det J(E_i)>0$, and hence $\operatorname{Re}\lambda(\delta_H)=0$.
Moreover,
\begin{align}
\left.
\frac{d\,\operatorname{Re}\lambda(\delta)}{d\delta}\right|_{\delta=\delta_H}
&=\frac{1}{2q_x(x_i,\delta_H)}
\left[r'(x_i)M'(x_i)+r(x_i)M''(x_i)-(p-1)^2\bigl(\beta N(x_i)-\delta_H\bigr)\frac{M'(x_i)}{M(x_i)}\right].
\label{eq:transversality}
\end{align}
Using
\[
\beta N(x_i)-\delta_H=\frac{r(x_i)M'(x_i)}{p-1},
\]
the transversality condition reduces to
\begin{equation}\label{eq:hopf_trans_condition}
r'(x_i)M'(x_i)+r(x_i)M''(x_i)-(p-1) r(x_i) \frac{\bigl(M'(x_i)\bigr)^2}{M(x_i)} \neq 0.
\end{equation}

Therefore, we obtain the following result.

\begin{theorem}\label{thm:hopf}
Let $0<p<1$ and let $E_i=(x_i,M(x_i))$ be an interior equilibrium
corresponding to $\delta=\delta_H$. If
\[
\beta N'(x_i)M(x_i)-(p-1)\bigl(\beta N(x_i)-\delta_H\bigr)M'(x_i)>0
\]
and
\[
r'(x_i)M'(x_i)+r(x_i)M''(x_i)-(p-1) r(x_i)\frac{\bigl(M'(x_i)\bigr)^2}{M(x_i)}
\neq0,
\]
then system~\eqref{eq:model} undergoes a generic Hopf bifurcation at $E_i$ as $\delta$ passes through $\delta_H$.
\end{theorem}
Having established the conditions for the occurrence of Hopf bifurcation, in the next subsection, we will determine the codimension of Hopf bifurcation.
\subsubsection{Codimension of Hopf bifurcation}
To determine the criticality of the Hopf bifurcation and identify the
possible occurrence of a generalized Hopf bifurcation, we transform the system into a form suitable for the computation of the Lyapunov coefficients~\cite{perko2013differential, caubergh2004hopf, lamontagne2008bifurcation}. Following the approach used for planar predator-prey
systems, we first translate the interior equilibrium to the origin and
then introduce a time rescaling and a near-identity change of coordinates.

Recall that the system can be written as
\begin{equation}\label{eq:system_GNR}
\begin{aligned}
\dot{x} &= r(x)\bigl(M(x)-y\bigr),\\
\dot{y} &= y\left(\beta N(x)-\delta-c y^{p-1}\right),
\end{aligned}
\end{equation}
where
\[
r(x)=\frac{x}{x+\sigma},\qquad
M(x)=\left(1-\frac{x}{\gamma}\right)(x+\sigma),
\qquad
N(x)=\frac{x+\xi}{x+\sigma}.
\]

Let \(E_i=(x_i,y_i)\) be an interior equilibrium. Then
\begin{equation}\label{eq:interior_eq_relations}
y_i=M(x_i),\qquad
\beta N(x_i)-\delta-c y_i^{p-1}=0.
\end{equation}
We first translate \(E_i\) to the origin by introducing
\[
x=x_i+x_1,\qquad y=y_i+y_1.
\]
Using \eqref{eq:interior_eq_relations}, system
\eqref{eq:system_GNR} becomes
\begin{equation}\label{eq:translated_system}
\begin{aligned}
\dot{x}_1&=r(x_i+x_1)\left[M(x_i+x_1)-M(x_i)-y_1\right],\\
\dot{y}_1&=(y_i+y_1)\left[\beta N(x_i+x_1)-\delta-c(y_i+y_1)^{p-1}\right].
\end{aligned}
\end{equation}

Since \(x_i>0\), we have \(r(x_i+x_1)>0\) in a sufficiently small
neighborhood of the origin. Hence, we may introduce the time rescaling
\begin{equation}\label{eq:time_rescaling}
d\tau=r(x_i+x_1)\,dt.
\end{equation}
This positive time rescaling preserves the orientation of trajectories
and, in particular, does not change the existence or stability of
periodic orbits. In the new time variable, system \eqref{eq:translated_system} takes the form
\begin{equation}\label{eq:rescaled_system}
\begin{aligned}
\dot{x}_1&=g(x_1)-y_1,\\
\dot{y}_1&=\frac{y_i+y_1}{r(x_i+x_1)}\left[\beta N(x_i+x_1)-\delta-c(y_i+y_1)^{p-1}\right],
\end{aligned}
\end{equation}
where
\begin{equation}\label{eq:def_g}
g(x_1):=M(x_i+x_1)-M(x_i).
\end{equation}

We next introduce the change of coordinates
\begin{equation}\label{eq:second_change}
x=x_1,\qquad y=y_1-g(x_1).
\end{equation}
So, system \eqref{eq:rescaled_system} is transformed into
\begin{equation}\label{eq:lienard_transformed}
\begin{aligned}
\dot{x}&=-y,\\
\dot{y}&=\frac{y_i+y+g(x)}{r(x_i+x)}\left[\beta N(x_i+x)-\delta-c\bigl(y_i+y+g(x)\bigr)^{p-1}\right]+g'(x)y.
\end{aligned}
\end{equation}

Expanding the right-hand side of \eqref{eq:lienard_transformed} in a Taylor series about the origin gives
\begin{equation}\label{eq:local_Hopf_expansion}
\begin{aligned}
\dot{x}&=-y,\\
\dot{y}&=a_{10}x+a_{01}y+a_{20}x^2+a_{11}xy+a_{02}y^2+a_{30}x^3+a_{21}x^2y+a_{12}xy^2+a_{03}y^3+O\!\left(|(x,y)|^4\right),
\end{aligned}
\end{equation}
where the coefficients \(a_{ij}\) are evaluated at the interior
equilibrium \(E_i\) and depend smoothly on the model parameters.

On the Hopf surface,
\[
\operatorname{tr}J(E_i)=0,
\]
and hence the coefficient of the linear \(y\)-term vanishes, $a_{01}=0$.

Thus, near a Hopf point, \eqref{eq:local_Hopf_expansion} reduces to
\begin{equation}\label{eq:Hopf_polynomial_form}
\begin{aligned}
\dot{x}&=-y,\\
\dot{y}&=a_{10}x+a_{20}x^2+a_{11}xy+a_{02}y^2+a_{30}x^3+a_{21}x^2y+a_{12}xy^2+a_{03}y^3+O\!\left(|(x,y)|^4\right).
\end{aligned}
\end{equation}
Since \(\det J(E_i)>0\) at a Hopf point, we have \(a_{10}>0\).
Therefore, the linearized system has eigenvalues
\[
\lambda_{1,2}=\pm i\sqrt{a_{10}},
\]
and the first Lyapunov coefficient can be computed from the nonlinear
coefficients in \eqref{eq:Hopf_polynomial_form} and is given by~\cite{perko2013differential, caubergh2004hopf, andronov1971theory}
\begin{equation}\label{eq:L1_general}
\begin{aligned}
L_1&=\frac{3a_{03}a_{10}^{2}-a_{11}(a_{02}a_{10}+a_{20})+a_{10}a_{21}}
{8a_{10}^{2}}\\
&=\frac{x_i^2 \mathcal{N}_{L_1}}{16\gamma^3(x_i+\sigma)^4a_{10}^2},
\end{aligned}
\end{equation}
where $\mathcal{N}_{L_1}$ is a quadratic polynomial in $\beta$ given by
\begin{equation}
    \mathcal{N}_{L_1}= A_{\mathrm{GH}}\beta^2+B_{\mathrm{GH}}\beta+C_{\mathrm{GH}}
\end{equation}
with
\begin{align*}
A_{\mathrm{GH}}=& -\gamma^2 (\sigma-\xi)^2 (p-2) (p-1) (\gamma-\sigma -2 x_i),\\
B_{\mathrm{GH}}=&-\gamma (\sigma-\xi) \left(\gamma^2 (2 (p-1) x_i-(p-2) \sigma )+2 K \left(p \sigma ^2+(5-4 p) x_i^2+\sigma  x_i\right)-(p-2) \sigma ^3\right.\\
& \left. + 8 (p-1) x_i^3+2 (2 p-1) \sigma x_i^2-2 (p-2) \sigma ^2 x_i\right),\\
C_{\mathrm{GH}}=& 2 x_i (\gamma-\sigma -2 x_i) \left(\gamma  \sigma  (\gamma -\sigma )+3 \sigma  x_i (\sigma -\gamma )+2 x_i^3+6 \sigma  x_i^2\right)\\
\end{align*}
From $\operatorname{tr}J=0$, we have $M'(x_i)=\frac{\gamma-2x_i-\sigma}{\gamma}<0$ which gives $\gamma<2x_i+\sigma$.
Throughout the analysis we assume
\begin{equation}\label{eq:basic_assumptions}
x_i>0,\qquad 0<x_i<\gamma<2x_i+\sigma,\qquad \sigma>0,\qquad \sigma-\xi>0,\qquad 0<p<1.
\end{equation}
$L_1=0$ reduces to $ \mathcal{N}_{L_1}=0$ which is a quadratic equation in $\beta$.

Its two algebraic solutions may be written as
\begin{equation}\label{eq:eta_GH_pm}
\beta_{\mathrm{GH}}^{\pm}=\frac{S_{\mathrm{GH}}\pm\sqrt{\Delta_{\mathrm{GH}}}}{2\gamma (\sigma-\xi)(p-2)(p-1)(\gamma-2x_i-\sigma)},
\end{equation}
where
\begin{align}
S_{\mathrm{GH}}={}&\gamma^2\left[2(p-1)x_i-(p-2)\sigma\right]+2\gamma\left[(5-4p)x_i^2+x_i\sigma+p\sigma^2\right]+
8(p-1)x_i^3\\
&\quad +2(2p-1)x_i^2\sigma-2(p-2)x_i\sigma^2-(p-2)\sigma^3,\nonumber \\
\Delta_{\mathrm{GH}}={}& S_{\mathrm{GH}}^2+8(p-2)(p-1)x_i(\gamma-2x_i-\sigma)^2Q_{\mathrm{GH}},
\label{eq:S_GH}
\end{align}
where
\begin{equation}\label{eq:Q_GH}
Q_{\mathrm{GH}}=2x_i^3+6x_i^2\sigma+\gamma(\gamma-\sigma)\sigma+3x_i\sigma(\sigma-\gamma).
\end{equation}

Under the assumptions \eqref{eq:basic_assumptions}, symbolic reduction
shows that
\begin{equation}\label{eq:Delta_positive}
\Delta_{\mathrm{GH}}>0.
\end{equation}
and hence $\beta_{\mathrm{GH}}^{\pm}$ are two distinct real solutions.

However, a generalized Hopf point must also satisfy the Hopf
nondegeneracy condition $\det J(E_i)>0$. Substitution of
$\beta_{\mathrm{GH}}^{\pm}$ into the determinant shows that
\[
\det J(E_i)>0
\]
has no admissible solution along the branch $\beta_{\mathrm{GH}}^{+}$, whereas it is satisfied on a nonempty
parameter region along the branch $\beta_{\mathrm{GH}}^{-}$.
Consequently, the admissible generalized Hopf branch is
\begin{equation}\label{eq:beta_GH}
\beta_{\mathrm{GH}}=\beta_{\mathrm{GH}}^{-}=\frac{-S_{\mathrm{GH}}-\sqrt{\Delta_{\mathrm{GH}}}}{2\gamma  (\sigma-\xi)(p-2)(p-1)(\gamma-2x_i-\sigma)}.
\end{equation}
Therefore, a generalized Hopf bifurcation may occur along
\eqref{eq:beta_GH}, provided that $\det J(E_i)>0$ and $L_2\neq0$.

Thus, $L_1=0$ determines an admissible parameter branch along which the Hopf bifurcation becomes degenerate. The occurrence of a codimension-2 generalized Hopf bifurcation further requires the nonvanishing of the second Lyapunov coefficient, $L_2$ and the nondegeneracy of the corresponding two-parameter unfolding. We summarize these in the following theorem.

\begin{theorem}\label{thm:gh}
    Let $E_i=(x_i,M(x_i))$ be an interior equilibrium satisfying $\operatorname{tr}J(E_i)=0$ and $\det J(E_i)>0$. Under the assumptions in \eqref{eq:basic_assumptions}, $L_1$ vanishes when $\beta=\beta_{\mathrm{GH}}^{-}$ given by \eqref{eq:beta_GH}. If $L_2\neq0$, and the corresponding two-parameter unfolding is nondegenerate, then the system undergoes a codimension-2 generalized Hopf bifurcation at $E_i$.
\end{theorem}

\begin{remark}
    The explicit computation of the second Lyapunov coefficient $L_2$ involves lengthy algebraic expressions and is not pursued here. Thus, the existence of a nondegenerate generalized Hopf bifurcation remains conditional on $L_2\neq 0$ and the regularity of the two-parameter unfolding.
\end{remark}

\subsection{Cusp Bifurcation}
\subsubsection{Existence of cusp bifurcation}\label{sec:cusp_exist}
A cusp bifurcation occurs when a saddle-node bifurcation becomes degenerate, corresponding to the coalescence of three equilibria. Recall that the interior equilibria of system \eqref{eq:model} are determined by the positive roots of
\[
q(x)=\beta N(x)-\delta-c M(x)^{p-1}=0,
\]
A cusp equilibrium $E_{**}=(x_{**},M(x_{**}))$ corresponds to a triple root of $q$ corresponding to Figure~\ref{fig:triple_root}, and hence satisfies
\[
q(x_{**})=q'(x_{**})=q''(x_{**})=0, \qquad q'''(x_{**})\neq0.
\]
From $q'(x^*)=0$, we obtain
\begin{equation}\label{eq:cusp_beta}
    \beta_{**}=c(p-1)\frac{M(x_{**})^{p-2}M'(x_{**})}{N'(x_{**})}.
\end{equation}
Substituting \eqref{eq:cusp_beta} into $q''(x_{**})=0$ gives
\begin{equation}\label{eq:cond2}
    M'(x_{**})\frac{N''(x_{**})}{N'(x_{**})}-(p-2)\frac{(M'(x_{**}))^2}{M(x_{**})}-M''(x_{**})=0.
\end{equation}
Using
\[
M'(x)=\frac{\gamma-\sigma-2x}{\gamma},
\qquad
M''(x)=-\frac{2}{\gamma},
\qquad
\frac{N''(x)}{N'(x)}=-\frac{2}{x+\sigma},
\]
\eqref{eq:cond2} reduces to
\begin{equation}\label{eq:cusp_condition}
p(\gamma-\sigma-2x_{**})^2=2(x_{**}+\sigma)^2.
\end{equation}
Since $0<p<1$ and $0<x_{**}<\gamma$, only the positive branch is
admissible, yielding
\begin{equation}\label{eq:cusp_gamma}
\gamma_{**}=\sigma+2x_{**}+\sqrt{\frac{2}{p}}(x_{**}+\sigma).
\end{equation}
Moreover, $M'(x_{**})>0$, and therefore $\beta_{**}>0$ requires
$N'(x_{**})<0$, or equivalently
\[
\xi>\sigma=1+\alpha\xi.
\]
Finally, $q(x_{**})=0$ gives
\begin{equation}\label{eq:cusp_delta}
\delta_{**}=\beta_{**}N(x_{**})-c M(x_{**})^{p-1}.
\end{equation}
To verify the cubic nondegeneracy, using \eqref{eq:cusp_beta} and \eqref{eq:cusp_gamma}, direct simplification gives
\begin{equation}\label{eq:cusp_qxxx}
q'''(x_{**})=c(p-1)M_{**}^{p-4}M_{**}'\frac{(x_{**}+\sigma)^2}{\gamma_{**}^2}\left[4(p+1)+6p\sqrt{\frac{2}{p}}\right].
\end{equation}
Since $0<p<1$ and $M_{**}'>0$, it follows that $q'''(x_{**})<0$. Hence $q'''(x_{**})\neq0$, and $x_{**}$ is a root of multiplicity exactly three. 

Also,
\[\det J(E_{**})=x_{**}\left(1-\frac{x_{**}}{\gamma_{**}}\right)q'(x_{**})=0,\]
while
\[
\operatorname{tr}J(E_{**})=r(x_{**})M'(x_{**})-(p-1)c M_{**}^{p-1}>0.
\]
Thus, $E_{**}$ has a simple zero eigenvalue. Therefore, provided
$\delta_{**}>0$, $E_{**}$ is a cusp candidate; the generic codimension-$2$ cusp follows upon verification of the two-parameter unfolding condition which is done in the following subsection.

\subsubsection{Two-parameter unfolding}
We now verify the two-parameter unfolding condition for cusp bifurcation. Although in Section~\ref{sec:cusp_exist}, the critical values $\gamma_{**}$, $\beta_{**}$, and $\delta_{**}$ are used to locate the cusp point, a cusp bifurcation is of codimension two, so we fix $\gamma=\gamma_{**}$ and take $\lambda=(\lambda_1,\lambda_2)=(\beta,\delta)$ as the unfolding parameters.

Since the interior equilibria are determined by $q(x;\beta,\delta)=0$, the unfolding matrix at the cusp is
\[
\mathcal{M} = \begin{pmatrix} q_\beta & q_\delta\\ q_{x\beta} & q_{x\delta}
\end{pmatrix}_{x=x_{**}}
=\begin{pmatrix}
N(x_{**}) & -1\\
N'(x_{**}) & 0
\end{pmatrix}.
\]
Hence
\[
\det(\mathcal{M}) = N'(x_{**}) = \frac{\sigma-\xi}{(x_{**}+\sigma)^2}.
\]
From the existence conditions established in Section~\ref{sec:cusp_exist}, $\xi>\sigma$, and therefore
\[
\det(\mathcal{M})\neq0,
\qquad
\operatorname{rank}(\mathcal{M})=2.
\]
Thus $(\beta,\delta)$ provides a generic two-parameter unfolding of
the cusp degeneracy.

More precisely, setting
\[
z=x-x_{**},\qquad
\mu_1=\beta-\beta_{**},\qquad
\mu_2=\delta-\delta_{**},
\]
the Taylor expansion of $q$ about the cusp gives
\[
q =N(x_{**})\mu_1-\mu_2+N'(x_{**})\mu_1z+\frac{q'''(x_{**})}{6}z^3
+O(z^4)
\]
Since $N'(x_{**})<0$ and $q'''(x_{**})<0$, a nonsingular change of
parameters and a rescaling of $z$ reduces the local equilibrium
equation to the standard cusp form
\[
\eta_1+\eta_2z- z^3=0.
\]
The following theorem summarizes the established result.
\begin{theorem}
\label{thm:cusp}
For $0<p<1$ and $\xi>\sigma$. Suppose that $\gamma=\gamma_{**}$, where $\gamma_{**}$ is given by \eqref{eq:cusp_gamma}, and $\beta_{**},\delta_{**}>0$ are given by \eqref{eq:cusp_beta} and \eqref{eq:cusp_delta}, respectively. Then system \eqref{eq:model_BT} undergoes a codimension-$2$ cusp bifurcation at the interior equilibrium $E_{**}=(x_{**},M(x_{**}))$, with $(\beta,\delta)$ as unfolding parameters.
\end{theorem}

\subsection{Saddle-node-transcritical bifurcation}
A saddle-node-transcritical (SNTC) bifurcation is a codimension-$2$ interaction that occurs when saddle-node and transcritical bifurcation curves meet in a two-parameter space. Such interactions have previously been reported in predator-prey systems, including Lotka-Volterra models with a Type-I functional response and models incorporating predator competition together with constant harvesting or migration terms \cite{broer2007dynamics,saputra2010saddle,saputra2015dynamical,parshad2023additional}. Depending on the eigenvalue configuration at the interaction point, SNTC bifurcations have been classified into the elliptic case, characterized by a single zero eigenvalue, and the saddle case, characterized by a double zero eigenvalue. Their numerical detection is also nontrivial, as standard continuation software such as MATCONT does not generally detect SNTC interactions directly; specialized numerical procedures have therefore been developed for their detection quite recently \cite{van2019automatic}.

For our system \eqref{eq:model_BT}, an SNTC bifurcation is observed at the intersection of the saddle-node curve $\mathcal C_{SN}$ and the transcritical curve $\mathcal C_T$ in the $(\delta,\xi)$-parameter plane; see Figure~\ref{fig:region_plots}. Along $\mathcal C_{SN}$, two interior equilibria coalesce, whereas along $\mathcal C_T$, an interior equilibrium collides with the positive predator-only equilibrium $E_P$. At their intersection, these two bifurcation mechanisms occur simultaneously.

In terms of the reduced equilibrium equation $q(x)=0$, the SNTC point satisfies
\[
q(0)=q_x(0)=0.
\]
For the parameter values used in Figure~\ref{fig:region_plots}, this occurs at
\[
(\delta_{\mathrm{SNTC}},\xi_{\mathrm{SNTC}})
\approx(1.77013618,1.25779729),
\]
with
\[
q(0)\approx-2.22\times10^{-16},\qquad
q_x(0)\approx1.39\times10^{-16},\qquad
q_{xx}(0)\approx0.11907\neq0.
\]
And, the Jacobian at the corresponding boundary equilibrium has eigenvalues
\[
\lambda_1=0, \qquad \lambda_2\approx0.30196>0,
\]
so that the SNTC is of single-zero type. The two-parameter unfolding is nondegenerate, with
\[
\det
\frac{\partial(q(0),q_x(0))}
     {\partial(\delta,\xi)}\approx1.65414\neq0.
\]
Hence, $(\delta_{\mathrm{SNTC}},\xi_{\mathrm{SNTC}})$ corresponds to a nondegenerate single-zero SNTC bifurcation and serves as a codimension-$2$ organizing center connecting the saddle-node and transcritical bifurcation curves.

\section{Numerical simulations}\label{sec:ns}
\subsection{Two-parameter bifurcation analysis}
\begin{figure}
    \centering
    \begin{subfigure}[t]{0.9\textwidth}
        \centering
        \includegraphics[width = 0.9\textwidth]{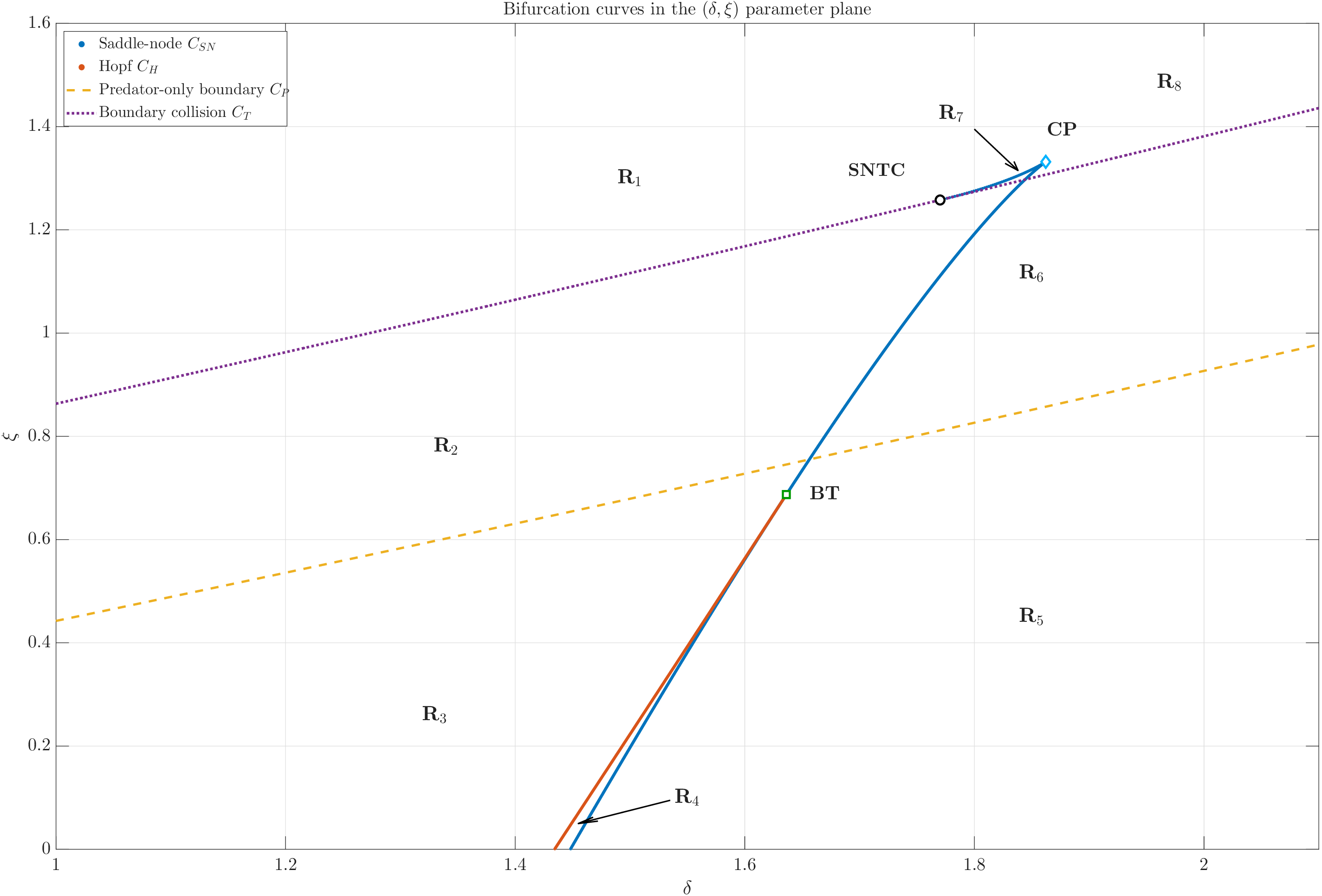}
        \caption{}
    \label{fig:region_plot1}
    \end{subfigure}\\
    ~ 
    \begin{subfigure}[t]{0.9\textwidth}
        \centering
        \includegraphics[width = 0.9\textwidth]{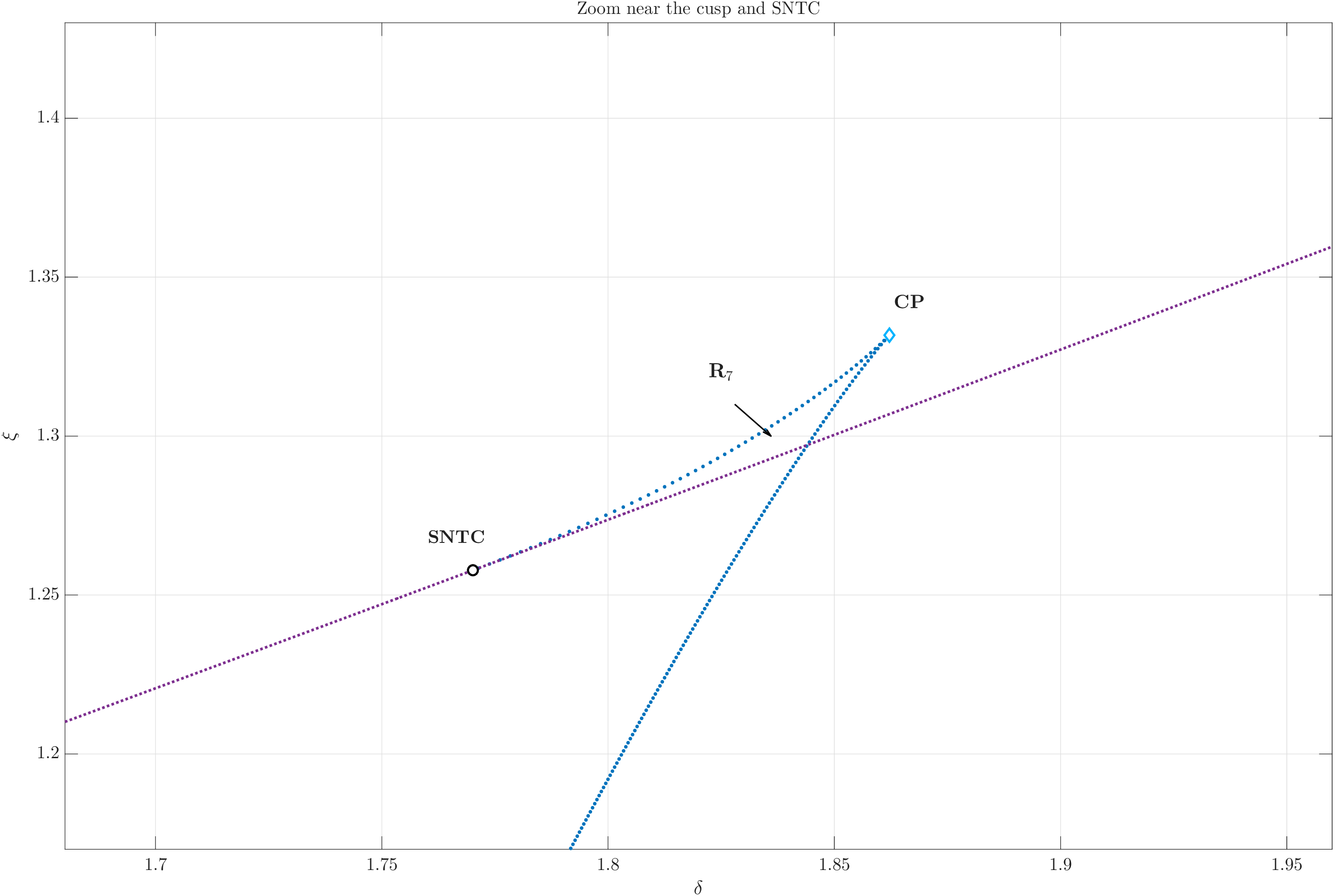}
        \caption{}
    \label{fig:region_plot2}
    \end{subfigure}
    \caption{Bifurcation diagram in the $(\delta,\xi)$ parameter plane: (a) complete diagram and (b) magnified view near the cusp and SNTC points. Parameters are $\gamma=10.97449$, $\alpha=0.10206$, $\beta=2.36197$, $c=0.90$, and $p=0.65$.}
    \label{fig:region_plots}
\end{figure}

Figure~\ref{fig:region_plots} summarizes the bifurcation structure of system~\eqref{eq:model_BT} in the $(\delta,\xi)$-parameter plane. The figure comprises of saddle-node curve $C_{SN}$, the Hopf curve $C_H$, the predator-only boundary $C_P$, and the boundary-collision curve $C_T$ partition the parameter plane into eight dynamically distinct regions, $R_1,\ldots,R_8$. Table~\ref{tab:regions}, describes the dynamics in each of these regions depending on the change in the quantity of AF given.

\begin{table}
\centering
\caption{Representative parameter values and equilibrium configurations
in the regions of the $(\delta,\xi)$-parameter plane.}
\label{tab:regions}
\renewcommand{\arraystretch}{1.25}
\begin{tabular}{|c| c| c| c| c| c|}
\hline
Region & No. of interior equilibria & Interior configuration & Positive boundary equilibrium\\
\hline
$R_1$ & $1$ & $S$ & $E_P$ exists\\

$R_2$ & $2$ & $UN+S$ & $E_P$ exists\\

$R_3$ & $2$ & $UF+S$ & No $E_P$\\

$R_4$ & $2$ & $SF+S$ & No $E_P$\\

$R_5$ & $0$ & None & No $E_P$\\

$R_6$ & $0$ & None & $E_P$ exists\\

$R_7$ & $3$ & $S+UN+S$ & $E_P$ exists\\

$R_8$ & $1$ & $S$ & $E_P$ exists\\
\hline
\end{tabular}

\medskip
\begin{flushleft}
\footnotesize
Here, $S$, $UN$, $UF$, and $SF$ denote a saddle,
unstable node, unstable focus, and stable focus, respectively. The positive predator-only boundary equilibrium is denoted by $E_P=(0,\tilde y)$. The boundary equilibria $E_0=(0,0)$ and
$E_\gamma=(\gamma,0)$ exist for all parameter values and are therefore not listed separately in the final column.
\end{flushleft}
\end{table}

The saddle-node curve $C_{SN}$ determines changes in the number of interior equilibria. In particular, crossing this curve results in the creation or annihilation of a pair of interior equilibria. The two branches of $C_{SN}$ meet at the cusp point $CP$, producing the narrow region $R_7$ in which three interior equilibria coexist. On either side of this region, the system possesses a single interior saddle, as observed in regions $R_1$ and $R_8$.

The Hopf curve $C_H$ separates regions with different stability of the non-saddle interior equilibrium. In $R_3$, the two interior equilibria consist of an unstable focus and a saddle and after crossing $C_H$, the unstable focus changes its stability through a Hopf bifurcation. The Hopf curve terminates at the Bogdanov-Takens point $BT$, where it meets the saddle-node curve $C_{SN}$.

%\textcolor{red}{Write about collision of $(0,y)$ and (x,y)}
The curve $C_P$ determines the existence of the positive predator-only boundary equilibrium $E_P=(0,\tilde y)$. Accordingly, $E_P$ exists in regions $R_1$, $R_2$, $R_6$, $R_7$, and $R_8$, whereas it does not exist in regions $R_3$, $R_4$, and $R_5$. Whenever it exists, $E_P$ is
an unstable node for $\tilde y<1+\alpha\xi$ and a saddle for $\tilde y>1+\alpha\xi$. The curve $C_T$, characterized by $\tilde y=1+\alpha\xi$, corresponds to a transcritical bifurcation at which an interior equilibrium coalesces with the predator-only equilibrium $E_P$. Across $C_T$, $E_P$ changes from an unstable node to a saddle.

The saddle-node curve $C_{SN}$ meets the boundary collision curve $C_T$ at a saddle-node-transcritical (SNTC) point,
\[
(\delta_{\mathrm{SNTC}},\xi_{\mathrm{SNTC}})
\approx(1.77013618,1.25779729),
\]
At this point, the reduced equilibrium equation satisfies
$q(0)=q_x(0)=0$ with $q_{xx}(0)\neq0$, while the corresponding boundary equilibrium has one zero and one nonzero eigenvalue. Thus, the SNTC point represents the collision of an interior saddle-node bifurcation with the
boundary equilibrium.

\subsection{Numerical simulations for FTE}
\begin{figure}
    \centering
    \includegraphics[width=\textwidth]{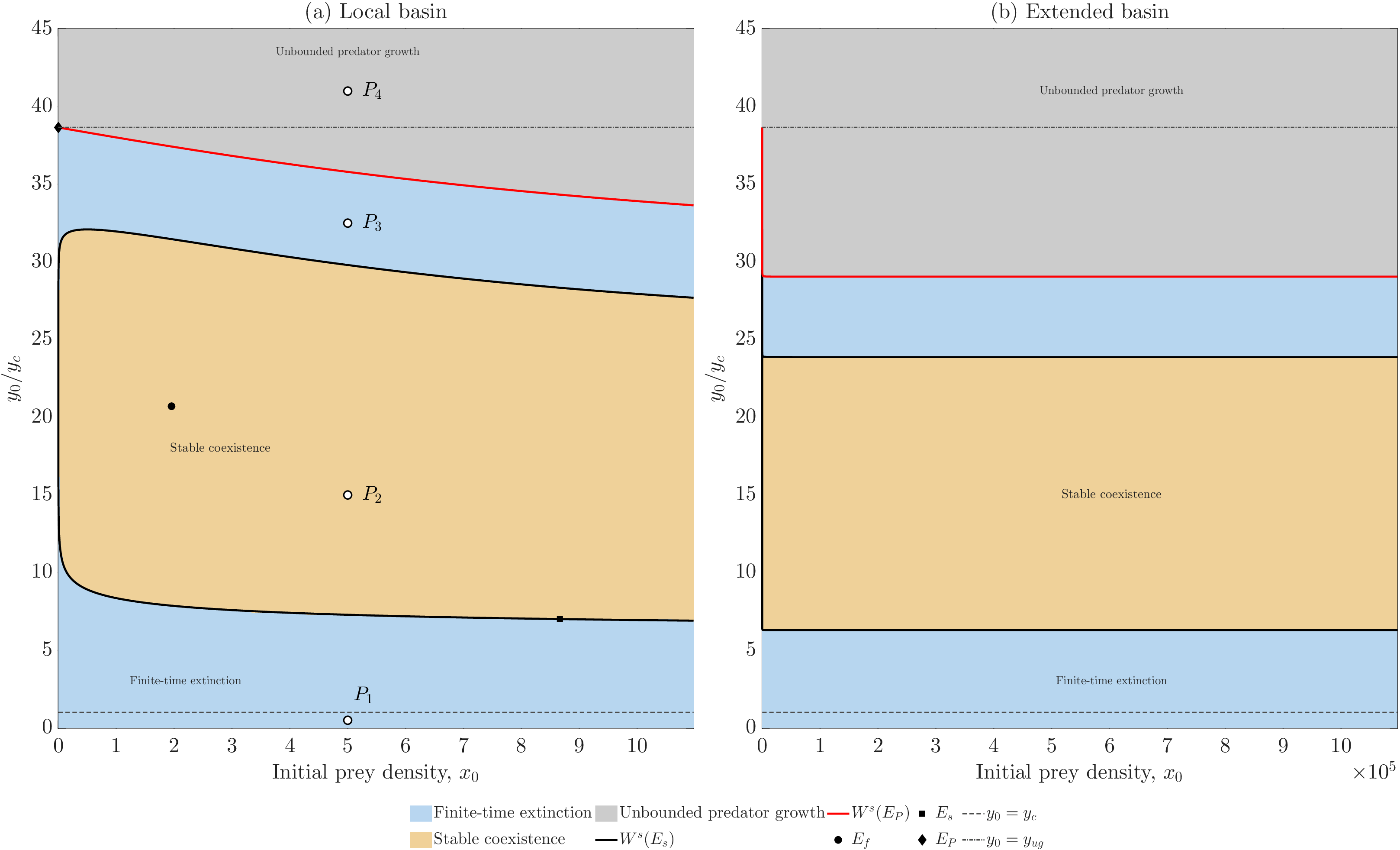}
    \caption{Basins of attraction in the $(x_0,y_0/y_c)$ plane. (a) Local basin showing the stable manifolds $W^s(E_s)$ and $W^s(E_P)$, with four selected initial conditions $P_1$-$P_4$ corresponding to the time series in Figure~\ref{fig:dynamics}. (b) Extended basin illustrating the persistence of the three dynamical outcomes for large initial prey densities. Parameters are $\gamma=10.97449$, $\alpha=2.74631$, $\xi=6.49166$, $\beta=0.35138$, $\delta=0.03238$, $c=0.29818$, and $p=0.65$.}
    \label{fig:FTE_basins}
\end{figure}

The sufficient condition obtained in Lemma~\ref{lem:FTE} guarantees finite-time extinction (FTE) of predator when the initial predator density is sufficiently small. However, FTE may also occur for initial densities above this threshold. To investigate how initial prey and predator densities affect the long-term dynamics of the system, we numerically find regions corresponding to FTE, stable coexistence and unbounded predator growth. Since the extinction threshold $y_c$ varies with the parameter values, we scale the initial predator density by $y_c$ and plot the results in the $(x_0,y_0/y_c)$-plane. With this scaling, the sufficient extinction threshold is given by the horizontal line $y_0/y_c=1$ (see Figure~\ref{fig:FTE_basins}).

Figure~\ref{fig:FTE_basins}(a) shows how initial population densities change the long-term dynamics. For the chosen parameter values, three distinct outcomes are observed: FTE, stable coexistence and unbounded predator growth. All sampled initial conditions below the threshold $y_0=y_c$ lead to FTE as predicted by Lemma~\ref{lem:FTE}. However, extinction also occurs for many initial conditions with $y_0>y_c$. Hence, the condition $y_0<y_c$ provides a sufficient but not necessary condition for FTE. Moreover, the extinction basin is not confined to small initial predator densities. Increasing $y_0$ can lead from FTE of predator to stable coexistence of populations followed by another region of FTE and eventually, unbounded predator growth in infinite time. This shows that increasing the initial predator population does not necessarily ensure its long-term persistence. 

To better understand the boundary separating these regions, we numerically compute the stable manifolds of the interior saddle equilibrium $E_s$ and the positive predator only equilibrium $E_P$. For the parameter values considered for this plot, $E_P$ is a saddle equilibrium. The stable manifold of a saddle equilibrium point consists of trajectories that approach the equilibrium as $t\to\infty$. To approximate its two branches numerically, we take small perturbations in opposite directions along the corresponding stable eigenvector and integrate the resulting trajectories backward in time. As shown in Figure~\ref{fig:FTE_basins}(a), $W^s(E_s)$ closely follows the numerically determined boundary separating the FTE basin from the stable-coexistence basin, whereas $W^s(E_P)$ follows the boundary between the upper FTE basin and the region of unbounded predator growth. This suggests that these stable manifolds act as separatrices, dividing the initial condition plane into regions with different long-term dynamical outcomes.

To see whether these outcomes persist for larger initial prey densities, Figure~\ref{fig:FTE_basins}(b) extends the range of $x_0$. It is observed that the three dynamical outcomes remain present.

\begin{figure}
    \centering
    \includegraphics[width=\textwidth, height=0.7\textwidth]{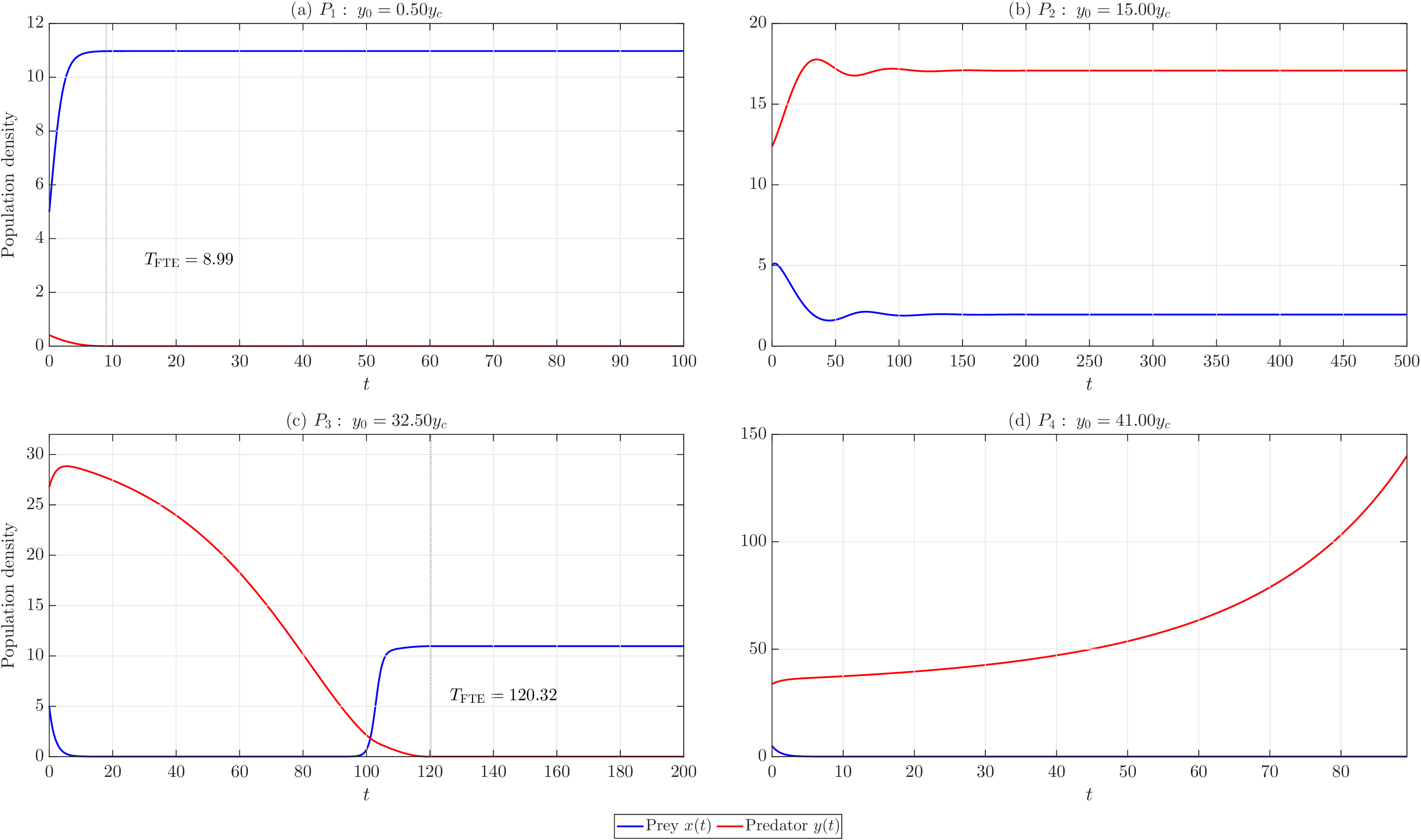}
    \caption{Time series corresponding to the four initial conditions marked in Figure~\ref{fig:FTE_basins}(a), with $x_0=5$. (a) $P_1$: $y_0=0.50y_c$, resulting in finite-time predator extinction (FTE) at $T_{\mathrm{FTE}}\approx 8.99$. (b) $P_2$: $y_0=15.00y_c$, resulting in stable coexistence. (c) $P_3$: $y_0=32.50y_c$, resulting in FTE at $T_{\mathrm{FTE}}\approx 120.32$, followed by prey recovery. (d) $P_4$: $y_0=41.00y_c$, illustrating unbounded predator growth while the prey approaches zero. Parameters are $\gamma=10.97449$, $\alpha=2.74631$, $\xi=6.49166$, $\beta=0.35138$, $\delta=0.03238$, $c=0.29818$, and $p=0.65$.}
    \label{fig:dynamics}
    \end{figure}

To further illustrate these dynamics, we select four initial conditions marked as $P_1$-$P_4$ in Figure~\ref{fig:FTE_basins}(a), and plot their corresponding time series in Figure~\ref{fig:dynamics}. For the four cases, the initial prey density is fixed at $x_0=5$, while the initial predator density is varied.

% \subsection{Unbounded predator growth}
% The unbounded-growth behavior described in Lemma~\ref{lem:unbounded_predator} is illustrated in
% Fig.~\ref{fig:unbounded_predator}. For the chosen parameter values, $k\approx1.95884>0$ and $y_{\mathrm{ug}}\approx4.07993$. Thus, the initial condition $y_0=5>y_{\mathrm{ug}}$ satisfies the sufficient
% condition of the lemma. The numerical solution shows that the predator population grows unboundedly, while the prey population approaches zero.
% \begin{figure}[h]
%     \centering
%     \includegraphics[width=0.78\linewidth]{figures/FTE/unbounded_predator_growth.eps}
%     \caption{Numerical illustration of the unbounded predator growth described in Lemma~\ref{lem:unbounded_predator}. The parameters used are $\gamma=10.97449$, $\alpha=0.10206$, $\xi=1.8$,
%     $\beta=2.36197$, $\delta=0.52285$, $c=1.78015$, and $p=0.65$,
%     with initial condition $(x_0,y_0)=(5,5)$.}
%     \label{fig:unbounded_predator}
% \end{figure}

\section{Dynamical comparison between different $p$ regimes}\label{sec:compare}
To understand how different $p$ regimes signifying different biological applications influence the dynamics of the predator-prey system, we compare the sublinear case $0<p<1$ with the classical AF model $(p=1,c=0)$ \cite{SP07} and the previously studied regimes $1<p<2$ and $p=2$ \cite{goyal2026additional, parshad2023additional, verma2026additional}. The comparison focuses on predator extinction, existence of pest extinction state, and the bifurcation structures arising under different forms of nonlinear predator loss.

Here, we present Table~\ref{tab:p_regime_qualitative} that summarizes the qualitative differences between these regimes. A key distinction is the possibility of FTE of predator when $0<p<1$. In this case, the per capita predator loss $cy^{p-1}$ increases as the predator density decreases, allowing the predator population to reach 0 in finite time, unlike the behavior we see in other regimes considered. Also, the pest extinction state (predator-only state) is always unstable whenver it exists in the sublinear case, whereas one can have different equilibrium configurations are in other regimes. 

Table~\ref{tab:p_regime_bifurcations} compares the bifurcation structures reported for different regimes. Codimension-1 bifurcations occur across several regimes, while higher-codimension bifurcations depend on the form of the nonlinear predator loss. For $0<p<1$, our analysis establishes the occurrence of a codimension-3 BT bifurcation, a codimension-2 generalized Hopf bifurcation, a codimension-2 cusp bifurcation, and a saddle-node--transcritical interaction. For $1<p\leq2$, previous studies reported higher-codimension BT, Hopf and homoclinic bifurcations, and the existence of multiple limit cycles. Thus, nonlinear predator loss can produce complex bifurcation structures in both sublinear and superlinear regimes, although their dynamical consequences differ.

\begin{figure}
    \centering
    \includegraphics[width=\textwidth, height=0.7\textwidth]{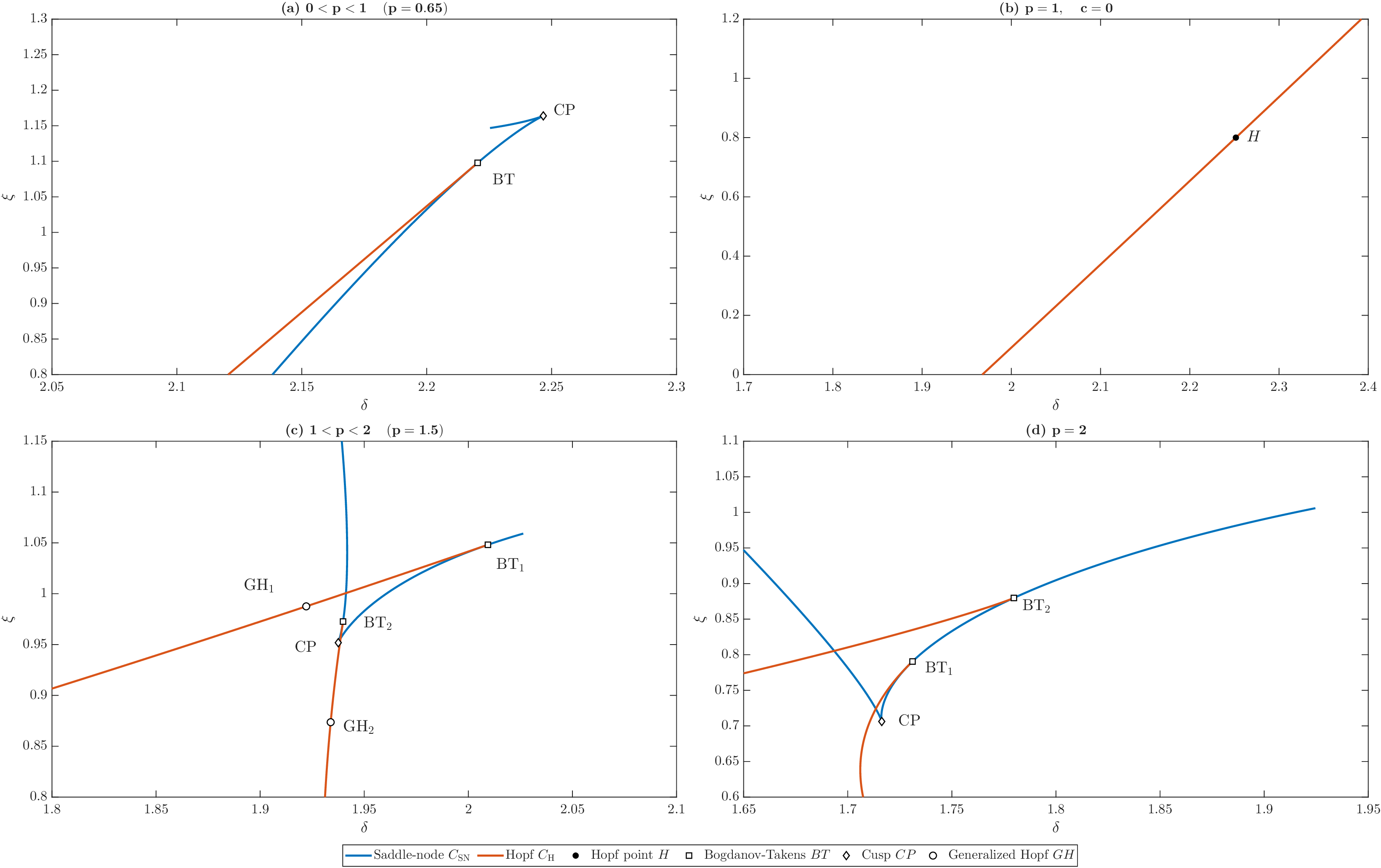}
    \caption{Bifurcation diagrams in the $(\delta,\xi)$-parameter plane illustrating the effect of the interference exponent $p$. (a) Sublinear interference, $0<p<1$, with $p=0.65$; (b) the classical AF model, $p=1$ and $c=0$; (c) sub-quadratic interference, $1<p<2$, with $p=1.5$; and (d) quadratic interference, $p=2$. For panels (a), (c), and (d), the remaining parameters are fixed at $\gamma=10.974492251988220$, $\alpha=0.1020627839$, $\beta=2.3619693707$, and $c=0.2074185$. The blue and orange curves denote the saddle-node curve $C_{\rm SN}$ and Hopf curve $C_{\rm H}$, respectively. Squares, diamonds, and open circles denote Bogdanov-Takens ($BT$), cusp ($CP$), and generalized Hopf ($GH$) points, respectively. The filled circle in panel (b) indicates a representative Hopf point $H$.}
\label{fig:p_regime_comparison}
\end{figure}

These differences were further illustrated with Figure~\ref{fig:p_regime_comparison} by comparing the bifurcation diagrams in the $(\delta,\xi)$-parameter plane for $p=0.65$, $p=1$, $p=1.5$, and $p=2$. The diagrams show how the saddle-node and Hopf curves, together with the associated special bifurcation points, vary across the different regimes. In particular, the sublinear case exhibits BT and cusp points, while the $1<p\leq2$ case display different arrangements of higher co-dimension bifurcation points. The classical AF model provides a reference case does not exhibit higher co-dimension bifurcations due to its interior equilibrium geometry. 

Overall, these results highlight the importance of the form of predator mortality or harvesting in understanding predator persistence and population dynamics in the presence of AF.

\begin{table}
\centering
\caption{Qualitative comparison across different $p$ regimes}
\label{tab:p_regime_qualitative}

\renewcommand{\arraystretch}{1.35}
\setlength{\tabcolsep}{5pt}
\small

\begin{tabular}{|C{2cm}|C{4.5cm}|C{2.5cm}|C{5.6cm}|}
\hline
\textbf{Regime}
& \textbf{Biological significance}
& \textbf{Predator FTE}
& \textbf{Predator-only state, $(0,\tilde{y})$} \\
\hline\hline

$p=1$, $c=0$
\cite{SP07}
& Classical AF model (no competition)
& No
& Possible (asymptotically) \\
\hline

$0<p<1$
& Enhanced predator mortality /
  sublinear harvesting
& Yes
& Unstable \\
\hline

$1<p<2$
\cite{goyal2026additional}
& Nonlinear competition / harvesting
& No
& Possible \\
\hline

$p=2$
\cite{goyal2026additional,parshad2023additional}
& Quadratic competition
& No
& Possible \\
\hline

\end{tabular}
\end{table}

\begin{table}
\centering
\caption{Comparison of bifurcation structures across different $p$ regimes}
\label{tab:p_regime_bifurcations}

\renewcommand{\arraystretch}{1.35}
\setlength{\tabcolsep}{5pt}
\small

\begin{tabular}{|C{2cm}|C{3.6cm}|C{3.7cm}|C{5.3cm}|}
\hline
\textbf{Regime}
& \textbf{Codim-1 bifurcations}
& \textbf{Codim-2 bifurcations}
& \textbf{Higher-codimension / other dynamics} \\
\hline\hline

$p=1$, $c=0$
\cite{SP07}
& \centering Saddle-node, transcritical, Hopf
& --
& Stable coexistence; stable limit cycle \\
\hline

$0<p<1$
& Saddle-node, transcritical, Hopf
& SNTC, degenerate Hopf, cusp
& BT (codim-3)\\
\hline

$1<p<2$
\cite{goyal2026additional,verma2026additional}
& \centering Saddle-node$^\dagger$, Hopf
& --
& Codim-4: Cusp-type BT;
Codim-3: focus-type BT, Hopf, homoclinic;
multiple limit cycle\\
\hline

$p=2$
\cite{goyal2026additional,parshad2023additional}
& Saddle-node, transcritical, Hopf
& SNTC, CTC, PTC
& Codim-4: Cusp-type BT;
Codim-3: focus-type BT, Hopf, homoclinic;
multiple limit cycle \\
\hline
\end{tabular}
\vspace{1mm}
\begin{minipage}{\textwidth}
\footnotesize
\textit{Note:} $^\dagger$ Implied by the local
unfolding of the Bogdanov--Takens bifurcation;
a separate saddle-node theorem is not established.
\end{minipage}
\end{table}

\section{Discussion \& Conclusion}\label{sec:discussion}
This work focused on investigating the dynamics of a predator-prey system incorporating additional food (AF) and sublinear mortality, which may account for enhanced predator mortality or sublinear harvesting, characterized by the exponent $0<p<1$. Motivated by the role of AF in the population dynamics and biological control and the significance of nonlinear predator loss, we considered a sublinear loss term of the form $-cy^p$. Our analysis demonstrates that sublinear mortality can substantially alter the qualitative dynamics of the system, giving rise to the existence of multiple interior equilibria, finite-time predator extinction, unbounded growth of predators as $t\to\infty$, and higher-codimension bifurcations. The non-Lipschitz nature of the mortality term at $y=0$ introduces dynamical features that are absent under conventional predator mortality.

We established the existence and stability properties of the boundary equilibria in Proposition~\ref{prop:boundary_stability} wherein we showed the prey-only equilibrium is asymptotically stable, whereas the predator-only equilibrium, whenever exists, can be an unstable node or a saddle depending on the parameter values. Further, in Theorem~\ref{thm:max_three_equilibria}, we derived that the system can admit at most three interior equilibria. Due to sublinear mortality, finite-time extinction (FTE) of predator is possible even in the presence of AF. In Lemma~\ref{lem:FTE}, a sufficient condition for FTE when the initial predator density is below a critical threshold, $y_c$, was derived. On the other hand, Lemma~\ref{lem:unbounded_predator} provides sufficient condition under which predator population blows up as $t\to\infty$. Remark~\ref{rem:prey_extinct} proves that unbounded predator growth is accompanied by asymptotic prey extinction. These analytical results demonstrate that contrasting long-term population outcomes, including predator extinction, coexistence, and unbounded predator growth, are possible due to sublinear mortality. These outcomes are illustrated numerically in Figure~\ref{fig:FTE_basins} where the stable manifolds of saddle equilibria separate the regions with different long-term behavior. Interestingly, increasing the initial predator density does not necessarily ensure persistence, and FTE may occur even beyond the sufficient extinction threshold, $y_c$. Some of the these dynamics are further presented with the time-series plots in Figure~\ref{fig:dynamics}.

We were also interested in investigating the complex bifurcation structures that may arise for this system. Building upon the equilibrium analysis, we established the existence of a codimension-$3$ cusp-type Bogdanov-Takens (BT) bifurcation in Theorem~\ref{thm:BT_existence} and correspondingly derived its normal form and unfolding in Theorem~\ref{thm:BT_unfolding}. The resulting unfolding reveals the complex dynamics organized by the nilpotent singularity around this BT point, like the occurrence of nearby saddle-node, Hopf, and homoclinic bifurcations. We further analyzed the degeneracy of the Hopf bifurcation and identified an admissible parameter branch where the first Lyapunov coefficient vanishes. We showed that a codimension-2 generalized Hopf (Bautin) bifurcation can occur if $L_2\neq0$ and the two-parameter unfolding is nondegenerate (Theorem~\ref{thm:gh}). In addition, the existence of a codimension-2 cusp bifurcation was found when the three interior equilibria coalesce and its two-parameter unfolding was derived. Interestingly, we also identified a nondegenerate codimension-$2$ saddle-node--transcritical (SNTC) interaction, at which an interior saddle-node bifurcation meets the predator-only boundary equilibrium. The SNTC point was characterized by a single-zero eigenvalue and a nondegenerate two-parameter unfolding. Moreover, these theoretical results were complemented with numerical simulations and two-parameter continuation using \textit{MATCONT}. The bifurcation diagram presented in Figure~\ref{fig:region_plots} identifies eight distinct regions with different equilibrium configurations and stability properties summarized in Table~\ref{tab:regions}. The saddle-node and Hopf curves, together with the BT, cusp, and SNTC points, illustrate the rich bifurcation structure arising from variations in predator mortality and AF availability.

There are several interesting applications of our proposed model \eqref{eq:model_BT}, to current pest management tactics. Most notably, pest management under global climate change, which is expected to increase rainfall, particularly in the North-Central United States \cite{lee2023ipcc}, which raises the risk of flooding in soybean crop fields. Current research shows that virulent soybean aphid biotypes, those that can survive on resistant soybean plants, have enhanced resilience to flooding \cite{lewis2025host}. This
may introduce a differential fitness response to flooding between aphid biotypes that could alter their
population dynamics. In particular, it may also cause the efficacy of bio-control predators, even those which are naturally occurring, such as the Asian Lady Beetle (\emph{H. axyridis}), to reduce. This reduction, may be even more pronounced on avirulent aphid biotypes, as well as on predator sub populations with a preference for avirulent biotypes - or even those predator sub populations that are adversely affected due to precipitation stress. All of the above situations fall within the modeling framework of \eqref{eq:model_BT}. A further interesting direction on the lines of the extinction risks discussed in \cite{vanselow2019very}, is to treat the $-cy^p$ term as $-c(x,t)y^p$, where the $c(x,t)$, could be fitted seasonally or spatially - herein one may allude to techniques of ``cutoff" type \cite{popovic2026burgers}. Such cutoff techniques can also be adopted to the patch and/or multiple bio control agent setting \cite{verma2026additional, verma2026t}. These directions are under current investigation.

In Section~\ref{sec:compare}, we compared the qualitative dynamics across different $p$ regimes. This covers our previous work where we consider generalized predator competition model \cite{goyal2026additional}, and the classical AF model \cite{SP07}. The comparison in Figure~\ref{fig:p_regime_comparison} illustrates differences in the arrangement of saddle-node and hopf bifurcation curves and the occurrence of higher-codimension bifurcation points. In contrast to generalized predator competition when $1<p\leq2$, the sublinear mortality regime admits FTE of predator as a direct consequence of the singular per capita loss rate near zero predator density. Thus, the nonlinear exponent plays a crucial role not only in the local bifurcation structure but also in the global persistence properties of the predator population. These qualitative differences have been summarized in Tables~\ref{tab:p_regime_qualitative}-\ref{tab:p_regime_bifurcations}. All in all \eqref{eq:model_BT}, is a \emph{minimal} AF mediated biological control model, in that it has features of Allee type dynamics, harvesting and enhanced predator mortality (such as due to extreme weather events), all encapsulated in one term. Depending on the choice of application, the dynamics of the chosen feature with AF can be investigated, so as to lay out an apt pest management tactic, particularly in the advent of climate change. 

\section{Funding}
KG, and RP gratefully acknowledge partial financial support from the USDA National Institute of Food and Agriculture through the Agricultural and Food Research Initiative (Grant No. 2023-67013-39157).

\section{Appendix}
\subsection{Coefficients $a_{i0}$, $a_{ij}$, $a_{0j}$, $b_{i0}$, $b_{i1}$ and $b_{i2}$}\label{sec:A1}
{\footnotesize
\begin{align*}
        a_{20}&=\zeta_1, \; a_{03}=-\frac{p(p-2) M'(x_*)}{6 M(x_*)^2 r(x_*)}, \; a_{04}=\frac{p(p-2) (p-3) M'(x_*)}{24 M(x_*)^3 r(x_*)^2},\; a_{13}=\frac{p (p(7-2 p)-6) M'(x_*)^2}{6 M(x_*)^3 r(x_*)},\\
        a_{30}&=\frac{1}{6M(x_*)^2}\Big[M(x_*)r(x_*)\Big\{3M'(x_*)r'(x_*)\Big(2M(x_*)M''(x_*)+p[M'(x_*)]^2\Big)\\
        &-3\beta M(x_*)M'(x_*)N''(x_*)-\beta M(x_*)^2N'''(x_*)\Big\}
        -3\beta M(x_*)^2r'(x_*)
        \Big(2M'(x_*)N'(x_*)\\
        &+M(x_*)N''(x_*)\Big)+p(x_*)^2\Big\{3M(x_*)^2[M''(x_*)]^2+p(p-2)[M'(x_*)]^4\Big\}\Big],\\
        a_{21}&=\frac{1}{2} \left(2 M''(x_*) r'(x_*)+M'(x_*) r''(x_*)-\frac{p r(x_*) M'(x_*) \left(M(x_*) M''(x_*)+(p-2) M'(x_*)^2\right)}{M(x_*)^2}+\beta  N''(x_*)\right), \\ 
        a_{12}&=\frac{3 p M'(x_*) r'(x_*)}{2 M(x_*) r(x_*)}+\frac{p(p-1) M'(x_*)^2}{M(x_*)^2}+\frac{r(x_*) r''(x_*)+r'(x_*)^2}{r(x_*)^2}, \\
        a_{40}&=-\frac{1}{96M(x_*)^3r(x_*)}\Big[36\beta M(x_*)^3[r'(x_*)]^2\Big(2M'(x_*)N'(x_*)+M(x_*)N''(x_*)\Big)\\
        &+12M(x_*)^2r(x_*)\Big\{4\beta p [M'(x_*)]^2r'(x_*)N'(x_*)
        -3p[M'(x_*)]^3[r'(x_*)]^2\\
        &+M(x_*)M'(x_*)\Big(-9M''(x_*)[r'(x_*)]^2+2\beta(p+3)r'(x_*)N''(x_*) + 4\beta r''(x_*)N'(x_*)\Big)\\
        &
        +2\beta M(x_*)^2 \Big(r''(x_*)N''(x_*)+r'(x_*)N'''(x_*)\Big) \Big\}+M(x_*)r(x_*)^2\Big\{4M(x_*)^2\Big(6M''(x_*)(\beta N''(x_*)\\
        &-3M''(x_*)r'(x_*))+\beta M(x_*)N^{(4)}(x_*)\Big)
        +6p [M'(x_*)]^3\Big(\beta pN'(x_*)-4M(x_*)r''(x_*)\Big) \\
        & -48p(p-1)[M'(x_*)]^4r'(x_*)+4M(x_*)^2M'(x_*)\Big(\beta(p+4)N'''(x_*)-6M''(x_*)r''(x_*)\Big)\\
        &+3pM(x_*)[M'(x_*)]^2\Big(\beta(p+4)N''(x_*)-12M''(x_*)r'(x_*)\Big)\Big\}
        \\
        &+pq(x_*)^3M'(x_*)\Big\{-24M(x_*)^2[M''(x_*)]^2+\big(p(28-11p)-24\big)[M'(x_*)]^4\\
        &+3(16-9p)M(x_*)[M'(x_*)]^2M''(x_*)\Big\}\Big],\\
        a_{31}&=\frac{1}{12M(x_*)^3r(x_*)}\Big[M(x_*)r(x_*)\Big\{2M(x_*)^2\Big(3M''(x_*)r''(x_*)+\beta N'''(x_*)\Big)\\
        &+M(x_*)M'(x_*)\Big(2M(x_*)r'''(x_*)+3\beta p\,N''(x_*)\Big)\\
        &\qquad
        +3pM(x_*)[M'(x_*)]^2r''(x_*)-6p(p-2)[M'(x_*)]^3r'(x_*)\Big\}+pr(x_*)^2[M'(x_*)]^2\Big\{3(4-3p)M(x_*)M''(x_*)\\
        &\qquad
        -(p-2)(5p-6)[M'(x_*)]^2\Big\}+6G(x_*)^3q'(x_*)\Big\{2M''(x_*)r'(x_*)+M'(x_*)r''(x_*)+\beta N''(x_*)\Big\}\Big], \\
        a_{22}&=\frac{1}{8G(x_*)^3q(x_*)^3}\Big[M(x_*)r(x_*)^2\Big\{pM'(x_*)\Big(4M(x_*)r''(x_*)-(7p+4)M'(x_*)r'(x_*)\Big) +4M(x_*)^2r'''(x_*)\Big\}\\
        &-12M(x_*)^3[r'(x_*)]^3
        -20 p M(x_*)^2r(x_*)M'(x_*)[r'(x_*)]^2\\
        &+2pr(x_*)^3M'(x_*)\Big\{(p-2)M(x_*)M''(x_*)+2(p-3)(p-1)[M'(x_*)]^2\Big\}\Big],\\ 
    \end{align*}}
    \subsection{Coefficients $b_{i0}$, $b_{i1}$ and $b_{i2}$}\label{sec:A2}
\begin{align*}
b_{20}&=a_{20}=\zeta_1, \quad b_{30}=a_{30}, \quad b_{21}=a_{21}, \quad b_{12}=a_{12}, \quad b_{40}=a_{40}, \quad b_{31}=a_{31}-3a_{20}a_{03},\quad b_{22}=a_{22}.
\end{align*}

\subsection{Coefficients in Theorem~\ref{thm:BT_unfolding}}\label{sec:A3}
\subsubsection*{Coefficients in system~\eqref{eq:unfolding_system1}}
{\footnotesize
\begin{align*}
c_{00}&=\frac{x_*^2\lambda_3}{\gamma_*^2},\quad
c_{10}=\frac{x_*(\gamma_*-2x_*-\sigma)}{\gamma_*(x_*+\sigma)}
+\frac{2x_*\lambda_3}{\gamma_*^2},\quad
c_{01}=-\frac{x_*}{x_*+\sigma},\quad
c_{20}=\frac{\lambda_3}{\gamma_*^2}
+\frac{(\gamma_*-x_*)\sigma}{\gamma_*(x_*+\sigma)^2}
-\frac{1}{\gamma_*},\\
c_{11}&=-\frac{\sigma}{(x_*+\sigma)^2},
c_{30}=-\frac{(\gamma_*-x_*)\sigma}{\gamma_*(x_*+\sigma)^3},\quad
c_{21}=\frac{\sigma}{(x_*+\sigma)^3},\quad
c_{40}=\frac{(\gamma_*-x_*)\sigma}{\gamma_*(x_*+\sigma)^4},\quad
c_{31}=-\frac{\sigma}{(x_*+\sigma)^4},\\
d_{00}
&=\frac{\gamma_*-x_*}{\gamma_*}
\left[\lambda_1(x_*+\xi)-\lambda_2(x_*+\sigma)\right],\quad
d_{10}=\frac{x_*(\gamma_*-2x_*-\sigma)^2}{\gamma_*^2(x_*+\sigma)},\quad
d_{11}
=-\frac{(\beta+\lambda_1)(\xi-\sigma)}
{(x_*+\sigma)^2},\\
d_{01}&=-\lambda_2
+\frac{\lambda_1(x_*+\xi)}{x_*+\sigma}
-\frac{x_*(\gamma_*-2x_*-\sigma)}{\gamma_*(x_*+\sigma)},\quad
d_{11}
=-\frac{(\beta+\lambda_1)(\xi-\sigma)}
{(x_*+\sigma)^2},\quad
d_{02}
=-\frac{p x_*(\gamma_*-2x_*-\sigma)}
{2(\gamma_*-x_*)(x_*+\sigma)^2},
\\
d_{20}
&=\frac{(\gamma_*-x_*)(\beta+\lambda_1)(\xi-\sigma)}
{\gamma_*(x_*+\sigma)^2},
d_{30}
=-\frac{(\gamma_*-x_*)(\beta+\lambda_1)(\xi-\sigma)}
{\gamma_*(x_*+\sigma)^3},\quad
d_{40}
=\frac{(\gamma_*-x_*)(\beta+\lambda_1)(\xi-\sigma)}
{\gamma_*(x_*+\sigma)^4},\\
d_{21}
&=\frac{(\beta+\lambda_1)(\xi-\sigma)}
{(x_*+\sigma)^3},\quad
d_{03}
=-\frac{\gamma_* p(p-2)x_*(\gamma_*-2x_*-\sigma)}
{6(\gamma_*-x_*)^2(x_*+\sigma)^3},\quad
d_{31}
=-\frac{(\beta+\lambda_1)(\xi-\sigma)}
{(x_*+\sigma)^4},\\
d_{04}
&=-\frac{\gamma_*^2p(p-2)(p-3)x_*(\gamma_*-2x_*-\sigma)}
{24(\gamma_*-x_*)^3(x_*+\sigma)^4}.
\end{align*}
}
\subsubsection*{Coefficients in system~\eqref{eq:unfolding_system2}}
{\begingroup
\footnotesize
\begin{align*}
e_{00}={}&
c_{01}d_{00}-c_{00}d_{01}
+\frac{c_{00}^{2}}{c_{01}}d_{02}
-\frac{c_{00}^{3}}{c_{01}^{2}}d_{03}
+\frac{c_{00}^{4}}{c_{01}^{3}}d_{04}, \quad
e_{01}={}
c_{10}+d_{01}
-\frac{c_{00}(c_{11}+2d_{02})}{c_{01}}
+\frac{3c_{00}^{2}}{c_{01}^{2}}d_{03}
-\frac{4c_{00}^{3}}{c_{01}^{3}}d_{04},
\\[1mm]
e_{02}={}&
\frac{c_{11}+d_{02}}{c_{01}}
-\frac{3c_{00}}{c_{01}^{2}}d_{03}
+\frac{6c_{00}^{2}}{c_{01}^{3}}d_{04},\quad
e_{03}={}
\frac{d_{03}}{c_{01}^{2}}
-\frac{4c_{00}}{c_{01}^{3}}d_{04},\quad
e_{04}={}
\frac{d_{04}}{c_{01}^{3}}, \quad
e_{13}={}
-\frac{2(c_{11}d_{03}+2c_{10}d_{04})}{c_{01}^{3}}
+\frac{12c_{00}c_{11}}{c_{01}^{4}}d_{04},\\
e_{10}={}&c_{11}d_{00}-c_{10}d_{01}+c_{01}d_{10}-c_{00}d_{11}+\frac{2c_{00}c_{10}}{c_{01}}d_{02}-\frac{c_{00}^{2}c_{11}}{c_{01}^{2}}d_{02}-\frac{3c_{00}^{2}c_{10}}{c_{01}^{2}}d_{03}\\
&+\frac{2c_{00}^{3}c_{11}}{c_{01}^{3}}d_{03}+\frac{4c_{00}^{3}c_{10}}{c_{01}^{3}}d_{04}-\frac{3c_{00}^{4}c_{11}}{c_{01}^{4}}d_{04},
\\[1mm]
e_{11}={}&2c_{20}+d_{11}-\frac{c_{10}c_{11}+2c_{00}c_{21}+2c_{10}d_{02}}{c_{01}}+\frac{c_{00}c_{11}^{2}+2c_{00}c_{11}d_{02}
+6c_{00}c_{10}d_{03}}{c_{01}^{2}}\\
&-\frac{6c_{00}^{2}c_{11}d_{03}+12c_{00}^{2}c_{10}d_{04}}{c_{01}^{3}}+\frac{12c_{00}^{3}c_{11}}{c_{01}^{4}}d_{04},
\\[1mm]
e_{12}={}&\frac{2c_{21}}{c_{01}}-\frac{c_{11}(c_{11}+d_{02})+3c_{10}d_{03}}{c_{01}^{2}}
+\frac{6c_{00}(c_{11}d_{03}+2c_{10}d_{04})}{c_{01}^{3}}
-\frac{18c_{00}^{2}c_{11}}{c_{01}^{4}}d_{04},
\\[1mm]
e_{20}={}&
c_{21}d_{00}-c_{20}d_{01}+c_{11}d_{10}-c_{10}d_{11}+c_{01}d_{20}-c_{00}d_{21}+\frac{c_{10}^{2}+2c_{00}c_{20}}{c_{01}}d_{02}
\\
&-\frac{
2c_{00}c_{10}c_{11}d_{02}
+c_{00}^{2}c_{21}d_{02}
+3c_{00}c_{10}^{2}d_{03}
+3c_{00}^{2}c_{20}d_{03}
}{c_{01}^{2}}\\
&+\frac{
c_{00}^{2}c_{11}^{2}d_{02}
+6c_{00}^{2}c_{10}c_{11}d_{03}
+2c_{00}^{3}c_{21}d_{03}
+6c_{00}^{2}c_{10}^{2}d_{04}
+4c_{00}^{3}c_{20}d_{04}
}{c_{01}^{3}}
\\
&-\frac{
3c_{00}^{3}c_{11}^{2}d_{03}
+12c_{00}^{3}c_{10}c_{11}d_{04}
+3c_{00}^{4}c_{21}d_{04}
}{c_{01}^{4}}
+\frac{6c_{00}^{4}c_{11}^{2}}{c_{01}^{5}}d_{04},
\\[1mm]
e_{21}={}&
3c_{30}+d_{21}
-\frac{
c_{11}c_{20}+2c_{10}c_{21}
+3c_{00}c_{31}+2c_{20}d_{02}
}{c_{01}}\\
&+\frac{
c_{10}c_{11}(c_{11}+2d_{02})
+3c_{10}^{2}d_{03}
+c_{00}\!\left(
3c_{11}c_{21}+2c_{21}d_{02}+6c_{20}d_{03}
\right)
}{c_{01}^{2}}
\\
&-\frac{
c_{00}\!\left(
c_{11}^{3}+2c_{11}^{2}d_{02}
+12c_{10}c_{11}d_{03}
+6c_{00}c_{21}d_{03}
+12(c_{10}^{2}+c_{00}c_{20})d_{04}
\right)
}{c_{01}^{3}}
\\
&+\frac{
3c_{00}^{2}\!\left(
3c_{11}^{2}d_{03}
+12c_{10}c_{11}d_{04}
+4c_{00}c_{21}d_{04}
\right)
}{c_{01}^{4}}
-\frac{24c_{00}^{3}c_{11}^{2}}{c_{01}^{5}}d_{04},
\\[1mm]
e_{22}={}&
\frac{3c_{31}}{c_{01}}
-\frac{
3c_{11}c_{21}+c_{21}d_{02}+3c_{20}d_{03}
}{c_{01}^{2}}
+\frac{
c_{11}^{3}+c_{11}^{2}d_{02}
+6c_{10}c_{11}d_{03}
+6c_{00}c_{21}d_{03}
+6(c_{10}^{2}+2c_{00}c_{20})d_{04}
}{c_{01}^{3}}
\\
&-\frac{
9c_{00}\!\left(
c_{11}^{2}d_{03}
+4c_{10}c_{11}d_{04}
+2c_{00}c_{21}d_{04}
\right)
}{c_{01}^{4}}
+\frac{36c_{00}^{2}c_{11}^{2}}{c_{01}^{5}}d_{04},\\
e_{30}={}&
c_{31}d_{00}-c_{30}d_{01}
+c_{21}d_{10}-c_{20}d_{11}
+c_{11}d_{20}-c_{10}d_{21}
+c_{01}d_{30}-c_{00}d_{31}
+\frac{2(c_{10}c_{20}+c_{00}c_{30})}{c_{01}}d_{02}
\\
&-\frac{
c_{10}^{2}c_{11}d_{02}
+c_{10}^{3}d_{03}
+2c_{00}c_{10}(c_{21}d_{02}+3c_{20}d_{03})
+c_{00}\!\left(
2c_{11}c_{20}d_{02}
+c_{00}c_{31}d_{02}
+3c_{00}c_{30}d_{03}
\right)
}{c_{01}^{2}}
\\
&+\frac{
2c_{00}c_{10}
\left(c_{11}^{2}d_{02}
+3c_{10}c_{11}d_{03}
+2c_{10}^{2}d_{04}\right)
}{c_{01}^{3}}
+\frac{
2c_{00}^{2}
\left(
c_{11}c_{21}d_{02}
+3c_{11}c_{20}d_{03}
+3c_{10}c_{21}d_{03}
+6c_{10}c_{20}d_{04}
\right)
}{c_{01}^{3}}
\\
&+\frac{
2c_{00}^{3}(c_{31}d_{03}+2c_{30}d_{04})
}{c_{01}^{3}}
-\frac{
c_{00}^{2}c_{11}
\left(c_{11}^{2}d_{02}
+9c_{10}c_{11}d_{03}
+6c_{00}c_{21}d_{03}\right)
}{c_{01}^{4}}+\frac{
4c_{00}^{3}c_{11}
\left(
c_{11}^{2}d_{03}
+6c_{10}c_{11}d_{04}
+3c_{00}c_{21}d_{04}
\right)
}{c_{01}^{5}}
\\
&-\frac{
3c_{00}^{2}
\left(
6c_{10}^{2}c_{11}
+4c_{00}c_{10}c_{21}
+c_{00}(4c_{11}c_{20}+c_{00}c_{31})
\right)d_{04}
}{c_{01}^{4}}
-\frac{10c_{00}^{4}c_{11}^{3}}{c_{01}^{6}}d_{04},
\\[1mm]
e_{31}={}&
4c_{40}+d_{31}
-\frac{
2c_{20}c_{21}+c_{11}c_{30}
+3c_{10}c_{31}+2c_{30}d_{02}
}{c_{01}}\\
&+\frac{
c_{11}^{2}c_{20}
+c_{11}(3c_{10}c_{21}+4c_{00}c_{31}+2c_{20}d_{02})
+2c_{10}(c_{21}d_{02}+3c_{20}d_{03})
}{c_{01}^{2}}
\\
&+\frac{
2c_{00}(c_{21}^{2}+c_{31}d_{02}+3c_{30}d_{03})
}{c_{01}^{2}}
-\frac{
c_{10}\!\left(
c_{11}^{3}+2c_{11}^{2}d_{02}
+6c_{10}c_{11}d_{03}
+4c_{10}^{2}d_{04}
\right)
}{c_{01}^{3}}-\frac{6c_{00}^{2}c_{31}}{c_{01}^{3}}d_{03}
-\frac{12c_{00}^{2}c_{30}}{c_{01}^{3}}d_{04}
\\
&-\frac{
4c_{00}\!\left(
c_{11}^{2}c_{21}
+c_{11}c_{21}d_{02}
+3c_{11}c_{20}d_{03}
+3c_{10}c_{21}d_{03}
+6c_{10}c_{20}d_{04}
\right)
}{c_{01}^{3}}\\
&+\frac{
c_{00}\!\left[
c_{11}^{3}(c_{11}+2d_{02})
+18c_{11}(c_{10}c_{11}+c_{00}c_{21})d_{03}
\right]
}{c_{01}^{4}}+\frac{40c_{00}^{3}c_{11}^{3}}{c_{01}^{6}}d_{04}
\\
&+\frac{
12c_{00}\!\left[
3c_{10}^{2}c_{11}
+3c_{00}c_{10}c_{21}
+c_{00}(3c_{11}c_{20}+c_{00}c_{31})
\right]d_{04}
}{c_{01}^{4}}
-\frac{
12c_{00}^{2}c_{11}
\left(
c_{11}^{2}d_{03}
+6c_{10}c_{11}d_{04}
+4c_{00}c_{21}d_{04}
\right)
}{c_{01}^{5}},\\
e_{40}={}&
-c_{40}d_{01}+c_{31}d_{10}-c_{30}d_{11}
+c_{21}d_{20}-c_{20}d_{21}
+c_{11}d_{30}-c_{10}d_{31}
+c_{01}d_{40}
+\frac{
c_{20}^{2}+2c_{10}c_{30}+2c_{00}c_{40}
}{c_{01}}d_{02}
\\
&-\frac{
\left[
c_{10}^{2}c_{21}
+2c_{00}(c_{20}c_{21}+c_{11}c_{30})
+2c_{10}(c_{11}c_{20}+c_{00}c_{31})
\right]d_{02}
}{c_{01}^{2}}
-\frac{
3\left[
c_{10}^{2}c_{20}
+2c_{00}c_{10}c_{30}
+c_{00}(c_{20}^{2}+c_{00}c_{40})
\right]d_{03}
}{c_{01}^{2}}
\\
&+\frac{
2c_{10}^{3}c_{11}d_{03}
+c_{10}^{4}d_{04}
+c_{10}^{2}c_{11}^{2}d_{02}
}{c_{01}^{3}}
+\frac{
2c_{00}c_{10}
\left(
2c_{11}c_{21}d_{02}
+6c_{11}c_{20}d_{03}
+3c_{00}c_{31}d_{03}
+6c_{00}c_{30}d_{04}
\right)
}{c_{01}^{3}}
\\
&+\frac{
6c_{00}c_{10}^{2}
(c_{21}d_{03}+2c_{20}d_{04})
}{c_{01}^{3}}
+\frac{
c_{00}\!\left(
2c_{11}^{2}c_{20}d_{02}
+2c_{00}c_{11}(c_{31}d_{02}+3c_{30}d_{03})
\right)
}{c_{01}^{3}}
\\
&+\frac{
c_{00}^{2}
\left(
c_{21}^{2}d_{02}
+6c_{20}c_{21}d_{03}
+6c_{20}^{2}d_{04}
+4c_{00}c_{40}d_{04}
\right)
}{c_{01}^{3}}
-\frac{
12c_{00}c_{10}^{3}c_{11}d_{04}
+9c_{00}c_{10}^{2}c_{11}^{2}d_{03}
+18c_{00}^{2}c_{10}^{2}c_{21}d_{04}
}{c_{01}^{4}}
\\
&-\frac{
2c_{00}c_{10}
\left(
c_{11}^{3}d_{02}
+6c_{00}^{2}c_{31}d_{04}
+9c_{00}c_{11}(c_{21}d_{03}+2c_{20}d_{04})
\right)
}{c_{01}^{4}}\\
&-\frac{
3c_{00}^{2}
\left[
c_{11}^{2}(c_{21}d_{02}+3c_{20}d_{03})
+c_{00}c_{21}(c_{21}d_{03}+4c_{20}d_{04})
+2c_{00}c_{11}(c_{31}d_{03}+2c_{30}d_{04})
\right]
}{c_{01}^{4}}
\\
&+\frac{
c_{00}^{2}c_{11}^{4}d_{02}
+12c_{00}^{2}c_{11}^{2}(c_{10}c_{11}+c_{00}c_{21})d_{03}
}{c_{01}^{5}}
+\frac{
6c_{00}^{2}
\left[
6c_{10}^{2}c_{11}^{2}
+8c_{00}c_{10}c_{11}c_{21}
+c_{00}(4c_{11}^{2}c_{20}
+c_{00}c_{21}^{2}
+2c_{00}c_{11}c_{31})
\right]d_{04}
}{c_{01}^{5}}
\\
&-\frac{
5c_{00}^{3}c_{11}^{2}
\left(
c_{11}^{2}d_{03}
+8c_{10}c_{11}d_{04}
+6c_{00}c_{21}d_{04}
\right)
}{c_{01}^{6}}
+\frac{15c_{00}^{4}c_{11}^{4}}{c_{01}^{7}}d_{04}.
\end{align*}
\endgroup
}
\subsubsection*{Functions in transformation~\eqref{eq:unfolding_transf_2}}
{\begingroup
\footnotesize
\begin{align*}
\Phi_2(x_2,y_2)
&=
\frac{e_{02}}{2}x_2^2
+\frac{e_{03}}{2}x_2^2y_2
+\frac{e_{04}}{2}x_2^2y_2^2
+\frac{1}{6}
\left(
2e_{02}^2
-2e_{01}e_{03}
-5e_{00}e_{04}
+e_{12}
\right)x_2^3+\frac{1}{6}
\left(
5e_{02}e_{03}
-4e_{01}e_{04}
+e_{13}
\right)x_2^3y_2
\\
&+\frac{1}{24}
\Big(
6e_{02}^3
+24e_{00}e_{03}^2
+6e_{01}^2e_{04}
-11e_{04}e_{10}
-4e_{03}e_{11}
+e_{02}\left(
-12e_{01}e_{03}
+6e_{00}e_{04}
+7e_{12}
\right)-2e_{01}e_{13}
+2e_{22}
\Big)x_2^4,\\
\Psi_2(x_2,y_2)
&=
e_{02}x_2y_2
+e_{03}x_2y_2^2
+e_{04}x_2y_2^3
+\frac{e_{00}e_{03}}{2}x_2^2
+\frac{1}{2}
\left(
2e_{02}^2
-e_{01}e_{03}
-3e_{00}e_{04}
+e_{12}
\right)x_2^2y_2
\\
&+\frac{1}{2}
\left(
5e_{02}e_{03}
-2e_{01}e_{04}
+e_{13}
\right)x_2^2y_2^2+\frac{1}{6}
\left(
3e_{03}e_{10}
+e_{00}
\left(
2e_{02}e_{03}
-4e_{01}e_{04}
+e_{13}
\right)
\right)x_2^3\\
&+\frac{1}{6}
\Big(
6e_{02}^3
+15e_{00}e_{03}^2
+2e_{01}^2e_{04}
-5e_{04}e_{10}
-e_{03}e_{11}-e_{01}e_{13}
+7e_{02}\left(-e_{01}e_{03}+e_{12}\right)
+2e_{22}
\Big)x_2^3y_2
\\
&+\frac{1}{12}
\Big(
9e_{00}^2e_{03}e_{04}
+7e_{02}e_{03}e_{10}
-8e_{01}e_{04}e_{10}
+2e_{10}e_{13}
\\
&\qquad
+e_{00}\left(
-10e_{02}^2e_{03}
+6e_{01}e_{03}^2+8e_{01}e_{02}e_{04}
-3e_{03}e_{12}
-2e_{02}e_{13}
\right)
+6e_{03}e_{20}
\Big)x_2^4.
\end{align*}
\subsubsection*{Coefficients in system~\eqref{eq:unfolding_system3}}
{\footnotesize
\begin{align*}
f_{00}={}&e_{00},
\quad
f_{01}={}e_{01},
\quad
f_{10}={}e_{10}-e_{00}e_{02},
\quad
f_{11}={}e_{11}-3e_{00}e_{03},
\\
f_{20}={}&
\frac{1}{2}\left[
-e_{02}e_{10}
+e_{00}\left(2e_{01}e_{03}+3e_{00}e_{04}-e_{12}\right)
+2e_{20}
\right],
\quad
f_{21}={}
\frac{1}{2}\left[
8e_{00}e_{01}e_{04}
-6e_{03}e_{10}
+e_{02}e_{11}
-3e_{00}e_{13}
+2e_{21}
\right],
\\[2mm]
f_{30}={}&
\frac{1}{6}\Big[
-e_{10}\left(e_{02}^2-4e_{01}e_{03}+2e_{12}\right)
+6e_{30}
-3e_{00}^2\left(5e_{03}^2+6e_{02}e_{04}\right)
\\
&\qquad\qquad
+e_{00}\left(
-6e_{01}^2e_{04}
+9e_{04}e_{10}
+4e_{03}e_{11}
-e_{02}e_{12}
+2e_{01}e_{13}
-2e_{22}
\right)
\Big],
\\[1mm]
f_{31}={}&
\frac{1}{6}\Big[
-9e_{02}e_{03}e_{10}
+24e_{01}e_{04}e_{10}
+2e_{02}^2e_{11}
-2e_{01}e_{03}e_{11}
+e_{11}e_{12}
-9e_{10}e_{13}
-18e_{03}e_{20}
+6e_{02}e_{21}
+6e_{31}
-63e_{00}^2e_{03}e_{04}
\\
&\qquad
+e_{00}\left(
78e_{02}^2e_{03}
-33e_{01}e_{03}^2
-60e_{01}e_{02}e_{04}
-5e_{04}e_{11}
+27e_{03}e_{12}
+21e_{02}e_{13}
\right)
\Big],
\\[2mm]
f_{40}={}&
\frac{1}{24}\Big[
54e_{00}^3e_{04}^2
-2e_{02}^3e_{10}
-18e_{01}^2e_{04}e_{10}
+9e_{04}e_{10}^2
+12e_{03}e_{10}e_{11}
+18e_{00}^2\left(
7e_{02}e_{03}^2
+2e_{02}^2e_{04}
+4e_{01}e_{03}e_{04}
-2e_{04}e_{12}
\right)
\\
&
+6e_{01}e_{10}e_{13}
-2e_{02}^2e_{20}
+8e_{01}e_{03}e_{20}
-4e_{12}e_{20}
-6e_{10}e_{22}
+e_{02}\left(
8e_{01}e_{03}e_{10}
-7e_{10}e_{12}
+12e_{30}
\right)
+24e_{40}
\\
&
+2e_{00}\Big(
12e_{02}^4
+12e_{01}^2e_{03}^2
+3e_{12}^2
+e_{02}^2\left(-36e_{01}e_{03}+22e_{12}\right)
-4e_{01}\left(2e_{04}e_{11}+3e_{03}e_{12}\right)
+2e_{11}e_{13}
-2e_{04}e_{20}
\\
&\qquad
+6e_{03}\left(-3e_{03}e_{10}+e_{21}\right)
+e_{02}\left(
24e_{01}^2e_{04}
-34e_{04}e_{10}
-3e_{03}e_{11}
-8e_{01}e_{13}
+8e_{22}
\right)
\Big)\Big].
\end{align*}}
\subsubsection*{Coefficients in system~\eqref{eq:unfolding_system4} and \eqref{eq:unfolding_system5}}
{\footnotesize
\begin{align*}
g_{00}={}&f_{00},
\quad
g_{01}={}f_{01},
\quad
g_{10}={}
f_{10}-\frac{f_{00}f_{30}}{2f_{20}},
\quad
g_{11}={}
f_{11}-\frac{f_{01}f_{30}}{2f_{20}},
\quad
g_{40}={}
\frac{
f_{10}\left(-15f_{30}^{3}
+16f_{20}f_{30}f_{40}\right)
}{
64f_{20}^{3}
},
\\[2mm]
g_{20}={}&
\frac{
80f_{20}^{3}
-60f_{10}f_{20}f_{30}
+45f_{00}f_{30}^{2}
-48f_{00}f_{20}f_{40}
}{
80f_{20}^{2}
},
\quad
g_{21}={}
f_{21}
-\frac{
3\left[
5f_{30}\left(4f_{11}f_{20}-3f_{01}f_{30}\right)
+16f_{01}f_{20}f_{40}
\right]
}{
80f_{20}^{2}
},
\\[2mm]
g_{30}={}&
\frac{
f_{10}\left(35f_{30}^{2}-32f_{20}f_{40}\right)
}{
40f_{20}^{2}
},
\quad
g_{31}={}
\frac{7f_{11}f_{30}^{2}}{8f_{20}^{2}}
+f_{31}
-\frac{
f_{21}f_{30}+\frac{4}{5}f_{11}f_{40}
}{
f_{20}
}.
\end{align*}
}

{\footnotesize
\begin{align*}
h_{00}
&=
\frac{g_{00}}{g_{20}}
-\frac{g_{10}^{2}}{4g_{20}^{2}},
\quad
h_{01}=
\frac{
-4g_{10}g_{11}g_{20}^{2}
+8g_{01}g_{20}^{3}
+2g_{10}^{2}g_{20}g_{21}
-g_{10}^{3}g_{31}
}
{8g_{20}^{7/2}},
\\[2mm]
h_{11}
&=
\frac{
4g_{11}g_{20}^{2}
-4g_{10}g_{20}g_{21}
+3g_{10}^{2}g_{31}
}
{4g_{20}^{5/2}},
\quad
h_{21}=
\frac{
2g_{20}g_{21}
-3g_{10}g_{31}
}
{2g_{20}^{3/2}},
\quad
h_{31}=
\frac{g_{31}}{\sqrt{g_{20}}}.
\end{align*}
}
\subsubsection*{Coefficients in system~\eqref{eq:unfolding_system6}}
{\footnotesize
\begin{align*}
J_{00} &= h_{00},
\quad
J_{01} = h_{01}-h_{00}h_{21},
\quad
J_{11} = h_{11},
\quad
J_{31} = h_{31}.
\end{align*}
}
\subsubsection*{Coefficients in system~\eqref{eq:unfolding_system7}}
{\footnotesize
\begin{align*}
\nu_0 = J_{00}J_{31}^{4/5},\quad
\nu_1 = J_{01}J_{31}^{1/5},\quad
\nu_2 = \frac{J_{11}}{J_{31}^{1/5}}.
\end{align*}
}

\subsubsection*{Functions in ~\eqref{eq:jac_cond}}
\begin{align*}
    \mathcal{F}_1&=\gamma_* (x_*+\sigma-(p-2) x_*)+(p-2) x_* (2x_*+\sigma)<0,\\
    \mathcal{F}_2&=\frac{x_*^2 \left(2 (x_*+\sigma) p (x_*-\gamma_*)+p (\gamma_*-x_*)^2+(x_*+\sigma)^2 (p-2)\right)}{\gamma_* (x_*+\sigma)}<0,\\
    \mathcal{P}(x_*)&=\sum_{j=0}^{7}\mathcal{P}_jx_*^j>0,
\end{align*}
with the polynomial coefficients being
\begin{align*}
\mathcal{P}_0={}&
\gamma_*^2 (\gamma_*-x_*-\sigma) (x_*+\sigma)^2
\left[-2(x_*+\sigma)^2+(\gamma_*-x_*-\sigma)^2p\right],\\[1mm]
\mathcal{P}_1={}&
-\gamma_*(x_*+\sigma)\Bigl(
2(x_*+\sigma)^2
\left(3\gamma_*^2-\gamma_*(x_*+\sigma)+3(x_*+\sigma)^2\right)
+(\gamma_*-x_*-\sigma)^4p^2\\
&\quad 
-(\gamma_*-x_*-\sigma)^2
\left(\gamma_*^2-5\gamma_*(x_*+\sigma)+5(x_*+\sigma)^2\right)p
\Bigr),\\[1mm]
\mathcal{P}_2={}&
12(x_*+\sigma)^3
\left(3\gamma_*^2-\gamma_*(x_*+\sigma)+(x_*+\sigma)^2\right)+(\gamma_*-x_*-\sigma)^4
\left(-3\gamma_*+11(x_*+\sigma)\right)p^2
+2(\gamma_*-x_*-\sigma)^5p^3\\
&\quad
+(x_*+\sigma)(\gamma_*-x_*-\sigma)
\left(-8\gamma_*^3+43\gamma_*^2(x_*+\sigma)
-44\gamma_*(x_*+\sigma)^2+20(x_*+\sigma)^3\right)p,\\[1mm]
\mathcal{P}_3={}&
6(x_*+\sigma)^3\left(-9\gamma_*+2(x_*+\sigma)\right)-10(\gamma_*-x_*-\sigma)^4p^3\\
&\quad
+(x_*+\sigma)
\left(22\gamma_*^3-121\gamma_*^2(x_*+\sigma)
+138\gamma_*(x_*+\sigma)^2-52(x_*+\sigma)^3\right)p\\
&\quad
+(\gamma_*-x_*-\sigma)^2
\left(15\gamma_*^2-52\gamma_*(x_*+\sigma)+43(x_*+\sigma)^2\right)p^2,\\[1mm]
\mathcal{P}_4={}&
2\left[
12(x_*+\sigma)^3
-2(x_*+\sigma)
\left(7\gamma_*^2-28\gamma_*(x_*+\sigma)+16(x_*+\sigma)^2\right)p\right.\\
&\qquad\left.
-(\gamma_*-x_*-\sigma)
\left(15\gamma_*^2-44\gamma_*(x_*+\sigma)+33(x_*+\sigma)^2\right)p^2
+10(\gamma_*-x_*-\sigma)^3p^3
\right],\\[1mm]
\mathcal{P}_5={}&
p\left[
10\gamma_*^2(3-2p)p
-2(x_*+\sigma)^2\left(18+5p(-5+2p)\right)+\gamma_*(x_*+\sigma)\left(17+p(-77+40p)\right)
\right],\\[1mm]
\mathcal{P}_6={}&
p\left[
5\gamma_*p(-3+2p)
+(x_*+\sigma)\left(-4+(19-10p)p\right)
\right], \qquad \mathcal{P}_7={}
(3-2p)p^2.
\end{align*}

%\newpage
\bibliographystyle{elsarticle-num}
\bibliography{references}

\end{document}